\documentclass[aps,prb,reprint,showpacs]{revtex4-1}
\usepackage{graphicx}
\def\bOmegaSO{\mathbf{\Omega}_\mathrm{SO}}
\def\OmegaSO{\Omega_\mathrm{SO}}

\def\bOmegaD{\mathbf{\Omega}_D}

\def\bOmegaR{\mathbf{\Omega}_R}
\def\eh{\textit{e-h}~}

\newcommand{\unitary}[1]{\hat{\mathbf{#1}}}

\begin{document}

%\preprint{v6, after PVS and AHM corrections}

\title{Electric control of spin transport in GaAs (111) quantum wells} %Title of paper



\author{A. Hern\'andez-M\'inguez}
\email[e-mail address: ]{alberto.h.minguez@pdi-berlin.de}
\author{K. Biermann}
\author{R. Hey}
\author{P. V. Santos}
%\thanks{}
\affiliation{Paul-Drude-Institut f\"ur Festk\"orperelektronik, Hausvogteiplatz 5-7, 10117 Berlin, Germany}


%\date{\today}

\begin{abstract}

We show by spatially and time-resolved photoluminescence that the  application of an electric field transverse to the plane of an intrinsic GaAs (111) quantum well (QW) allows the transport of photogenerated electron spins polarized along the direction perpendicular to the QW plane over distances exceeding 10~$\mu$m. We attribute the long spin transport lengths to the compensation of the in-plane effective magnetic field related to the intrinsic spin-orbit (SO) interaction by means of the electrically generated SO-field. Away from SO-compensation, the precession of the spin vector around the SO-field decreases the out-of-plane polarization of the spin ensemble as the electrons move away from the laser generation spot. The results are reproduced by a model for two-dimensional drift-diffusion of spin polarized charge carriers under weak SO-interaction.

\end{abstract}


\maketitle %\maketitle must follow title, authors, abstract and \pacs

%\section{}
%\label{}
%\subsection{}
%\subsubsection{}

\section{Introduction}

The manipulation of electron spins in semiconductors has attracted much interest during the last years due to potential applications for quantum  information processing.\cite{Awschalom2002,Zutic_ROMP76_323_04} Devices based on semiconductor spins require efficient techniques for the generation, storage, transport, and detection of the spin polarization, as well as interaction mechanisms for the manipulation of the spin vector in times shorter than the characteristic spin lifetime. One of the main challenges towards this goal in III-V semiconductor structures is the suppression of decoherence processes associated with the spin-orbit (SO) interaction, which reduces the lifetime of the electron spin polarization to values typically below one nanosecond and thus severely restricts its application in spintronic devices. 

The SO-interaction describes the coupling between the spin vector and the varying lattice potential acting on a moving electron. In the electron reference frame, the SO-interaction translates into an effective magnetic field, $\bOmegaSO$, which can lead to spin dephasing in an electron ensemble. Since $\bOmegaSO$ depends on the electron's wave vector, $\mathbf{k}$, its magnitude and direction changes after each electron scattering event. As a result, electron spins with the same initial polarization but moving with different wave vectors will precess around different axes with different Larmor frequencies in-between two consecutive scattering events, leading to the well-known Dyakonov-Perel' (DP) spin dephasing mechanism.\cite{Dyakonov_SPSS13_3023_72,Dyakonov_SPS20_110_86} Improvement of the electron spin lifetime in these materials demands, therefore, the control of spin dephasing related to the SO-interaction.

In GaAs-based quantum well (QW) structures, the SO-interaction is governed by two major contributions. The first one is associated with the bulk inversion asymmetry (BIA) of the zinc-blende lattice. This contribution, known as the Dresselhaus hamiltonian,\cite{Dresselhaus55a} $\mathcal{H}_D$, is expressed as:

\begin{equation}\label{Eq_BIA}
\mathcal{H}_D=\frac{\hat{\sigma}}{2}\hbar\bOmegaD(\mathbf{k})=\frac{\gamma}{2}\sum_i^{x,y,z}\sigma_ik_i\left(k^2_{i+1}-k^2_{i+2}\right),
\end{equation}

\noindent where $\hbar$ is the reduced Planck constant, $\hat{\sigma}=(\sigma_x,\sigma_y,\sigma_z)$ are the Pauli matrices, $k_i$ are the wave-vector components along $\langle100\rangle$, and $\gamma$ is the Dresselhaus spin-splitting constant of the material (we follow here the convention introduced by Cardona\cite{Christensen_SSC51_491_84,Cardona_PRB38_1806_88} and Eppenga\cite{Eppenga_PRB37_10923_88}).

The second important contribution to the SO-interaction arises from the structural inversion asymmetry (SIA) introduced by an external field. In most cases, the SIA is generated by an electric field, $\mathbf{E}$, leading to the Rashba Hamiltonian, $\mathcal{H}_R$:\cite{Bychkov1984}

\begin{equation}\label{Eq_SIA}
\mathcal{H}_R=\frac{\hat{\sigma}}{2}\hbar\bOmegaR(\mathbf{k},\mathbf{E})=r_{41}\hat{\sigma}\cdot\mathbf{k}\times\mathbf{E},
\end{equation}

\noindent where $r_{41}$ is the Rashba coefficient.\cite{Winkler03a} Since the strength of $\bOmegaR$ depends on $\mathbf{E}$, the Rashba contribution provides a powerful mechanism for the electric manipulation of the spin vector.

The impact of the SO-interaction on the electron spin dynamics depends on the crystallographic growth direction of the QW.\cite{Dyakonov_SPS20_110_86} In the last years, GaAs QWs grown along the $[111]$ direction have attracted an increasing interest due to the special symmetry of the Dresselhaus and Rashba effective magnetic fields in this case. In the reference frame defined by the axes $\unitary{x}=\frac{1}{\sqrt{6}}(-1,-1,2)$, $\unitary{y}=\frac{1}{\sqrt{2}}(1,-1,0)$, $\unitary{z}=\frac{1}{\sqrt{3}}(1,1,1)$, and for a QW subjected to a transverse electric field $\mathbf{E}=E_z\unitary{z}$, these two contributions can be expressed as:\cite{Cartoixa05a}

\begin{eqnarray}\label{Eq_BIA111}
\hbar\bOmegaD(\mathbf{k})=\frac{2\gamma}{\sqrt{3}}\left(\langle k_z^2\rangle -\frac{1}{4}k_{\parallel}^2\right)\left[\begin{array}{c}k_y \\ -k_x \\ 0\end{array}\right] \nonumber \\
+ \frac{\gamma}{\sqrt{6}}k_y(k_y^2-3k_x^2) \left[\begin{array}{c} 0 \\ 0 \\ 1\end{array}\right]
\end{eqnarray}

\begin{equation}\label{Eq_SIA111}
\hbar\bOmegaR(\mathbf{k},E_z)=2E_zr_{41}\left[\begin{array}{c}k_y \\ -k_x \\0 \end{array}\right].
\end{equation}

\noindent Here, $\langle k^2_z \rangle=\left(\pi/d_\mathrm{eff}\right)^2$ is the averaged squared wave vector along $z$ determined by the spatial extension of the electron wave function, ${d_\mathrm{eff}}$, and $k_\parallel=(k_x^2+k_y^2)^{1/2}$ is the in-plane wave vector amplitude. At low temperature and electron populations, quadratic terms in $k_x$ and $k_y$ can be neglected, so that both $\bOmegaD$ and $\bOmegaR$ lie in the QW plane and have exactly the same symmetry. As a result, by adjusting $E_z$ to satisfy the condition:

\begin{equation}\label{Eq_SOcomp}
\bOmegaD(\mathbf{k})+\bOmegaR(\mathbf{k},E_z^{c})=0,
\end{equation}

\noindent the SO-interaction becomes simultaneously suppressed for all values of $\mathbf{k}$ at a well-defined compensation electric field, $E_z^c$, leading to long electron spin lifetimes. This compensation mechanism was originally proposed in the theoretical works by Cartoix\`a\cite{Cartoixa05a} and Vurgaftman.\cite{Vurgaftman_JAP97_53707_05} Recent experiments have provided evidence for the electric enhancement of the spin lifetime,\cite{Balocchi2011,PVS257} as well as for the transition from a BIA-dominated to a SIA-dominated spin dephasing with increasing electric field.\cite{PVS261} The effects of the non-linear \textit{k} terms in $\bOmegaSO$ on the compensation mechanism have also been addressed.\cite{Sun_a__10,Wang_APL102_242408_13,Balocchi_NJ15_95016_13,PVS272}

A further important requirement for the efficient use of the spin degree of freedom in semiconductor devices is that carriers must be able to move from one point of the semiconductor to another one without losing their spin polarization. This requires not only long spin decoherence times but also good spin transport properties. Studies of spin diffusion in GaAs have been performed in the last years in bulk,\cite{JKDA99a, Crooker_Science_309_2191_05, Yu_APL94_202109_09, Quast_PRB79_245207_09, PhysRevB.87.205203, Weber_JAP109_106101_11, PhysRevB.88.195202} as well as in intrinsic,\cite{Cameron_PRL76_4793_96, Eldridge_PRB77_125344_08, Zhao_PRB79_115321_09, Hu_NRL6_149_11,PVS276} and n-doped QWs\cite{Weber_N437_1330_05, Carter_PRL97_136602_06, Volkl_PRB83_241306_11, Volkl_PRB89_75424_14, Altmann_PRB90_201306_14, Kohda_APL107_172402_15, Altmann_PRB92_235304_15} grown along different orientations. Although the spin transport properties were initially supposed to be the same as those of the charge, several results showed that this is not always true: while the charge transport in a carrier ensemble is not affected by carrier-carrier elastic scattering, this kind of interaction can play an important role in the case of spin transport. An example of this different behavior between carrier and spin transport is the spin Coulomb drag mechanism observed during unipolar spin diffusion in n-doped QWs.\cite{amico01a,Weber_N437_1330_05}

It has been recently shown that the spin diffusion length in GaAs (111) QWs can be efficiently controlled by a transverse electric field induced by a top electric gate.\cite{Wang_NC4_2372_13} Of special importance for future applications is the relation between the spin and carrier transport in these structures. In this contribution, we report experimental studies of both charge and spin transport in an intrinsic GaAs (111) QW under the effect of vertical electric fields. In our experiments, a cloud of out-of-plane spin polarized electron-hole (\textit{e-h}) pairs is optically generated by a tightly focused laser beam, and the expansion of this cloud under a transverse electric field is studied by spatially and time-resolved $\mu$-photoluminescence (PL). We show that the transport properties of out-of-plane electron spins depend on the applied electric field. We propose an explanation based on the precession of the spin vector around a weak in-plane SO-field during the radial expansion of the electron-hole cloud, finding a good agreement between simulation and experimental data. At SO-compensation, the suppressed precession of the spin vector around the SO-field allows the transport of the out-of-plane spin polarized electrons over distances exceeding $10~\mu$m. Under these conditions, the decay of the electron population in a few nanoseconds due to electron-hole recombination is the main mechanism limiting the spin transport distance.

This manuscript is organized as follows: in Section~\ref{Sec_setup} we introduce the sample design and the experimental techniques for the optical injection and detection of spin-polarized carriers. Section~\ref{subsec_carrier} shows the experimental results of charge transport obtained from both time-integrated and time-resolved experiments. Section~\ref{subsec_spin} shows the corresponding results for the case of spin transport, followed by a discussion of a spin drift-diffusion model in Section~\ref{Sec_discussion}. Finally, Section~\ref{Sec_summary} summarizes the main points of our work.

\section{Experimental details}\label{Sec_setup}

%*******************************************************************************
% Figure 1
\begin{figure}
\includegraphics[width=\columnwidth]{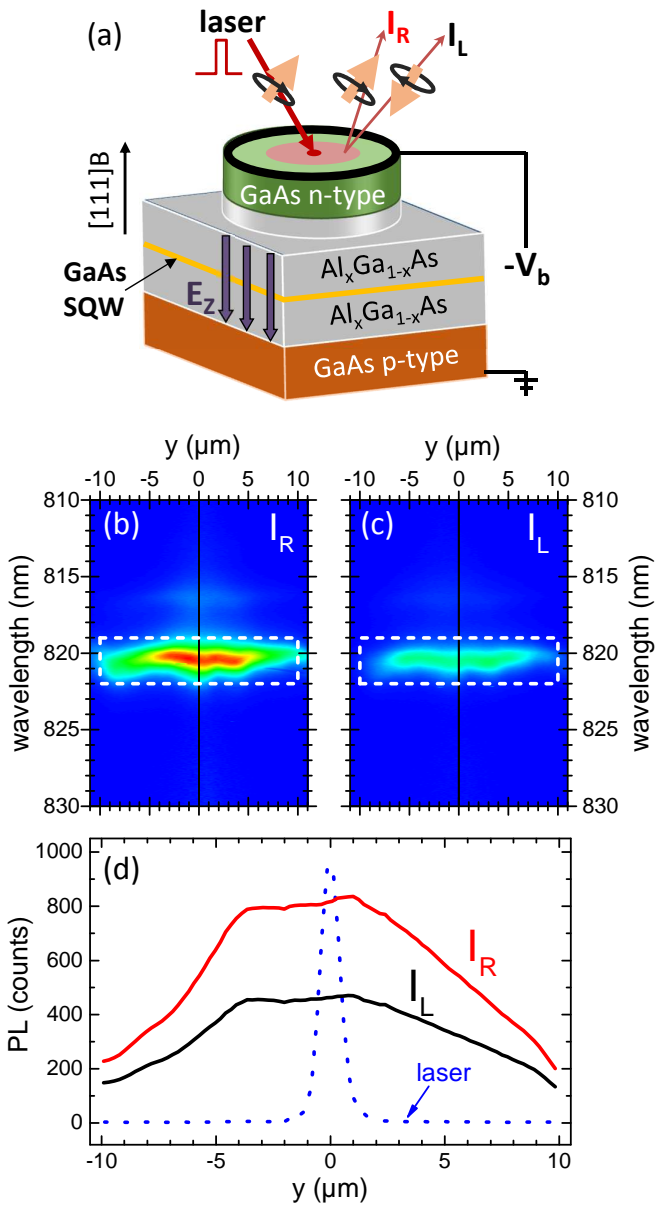}
\caption{(a) GaAs single quantum well (SQW) embedded in a \textit{n-i-p} structure grown along [111]B. The bias voltage, $V_b$, applied between the top $n$ region and the $p$-doped substrate generates the electric field, $E_z$, for the control of the SO-interaction. Spin polarized carriers are optically generated by a tightly focused pulsed laser. The carriers spread and recombine emitting photoluminescence (PL) away from the excitation spot. Panels (b) and (c) show the images of the time-integrated right ($I_R$) and left ($I_L$) circular PL components with wavelength (vertical scale) and position sensitivity along $y\parallel[1\overline{1}0]$ (horizontal scale) with respect to the generation point, $y=0$. The dashed boxes mark the PL of the \textit{e-hh} transition. (d) Intensity profiles of $I_R$ and $I_L$ emitted by the \textit{e-hh} transition along $y$. The dotted line shows the spatial distribution of the tightly focused laser beam, with a full width at half maximum, $\phi_L=1$~$\mu$m. The experiment was done at 30~K and $V_b=-1.5$~V.\label{setup}}
\end{figure}
%*******************************************************************************

The experiments reported here were carried out in an undoped single GaAs/AlGaAs QW with a thickness of 25~nm  embedded in the intrinsic region of a 
\textit{p-i-n} diode structure, cf.~Fig.~\ref{setup}(a). The sample was grown by molecular beam epitaxy on a $p$-doped GaAs(111)B substrate tilted by a small angle ($\delta\theta=2^\circ$) towards the $x$-direction (further details about the sample growth process can be found at Ref.~\onlinecite{PVS272}). The bias voltage, $V_b$, applied between an Al Schottky contact on top of the $n$-doped layer and the $p$-doped substrate, generates the vertical electric field $E_z$ required for the control of the SO-interaction. From previous measurements in a similar sample, we expect to reach $E_z^c$ at a reverse voltage bias $V_b\approx-1.5~$V.\cite{PVS261, PVS272} To confine the electric field along the $z$-direction, we processed the top doped layer and part of the top undoped (Al,Ga)As spacer into mesa structures of 300~$\mu$m diameter defined by wet chemical etching.

The spectroscopic studies were carried out with the sample in a cold finger cryostat at a temperature $T=30-40$~K equipped with a window for optical access and electric feed-throughs for the application of bias voltages. We optically excited \eh pairs with spin vector pointing perpendicular to the QW plane by using a circularly polarized, pulsed laser beam with wavelength of 756~nm, pulse width of 150 ps, and repetition rate of 40~MHz. The beam was spatially filtered and tightly focused onto a spot with a diameter (full width at half maximum)  $\phi_L=1~\mu$m by a 50$\times$ microscope objective. The carrier cloud generated by the laser beam spreads outwards leading to a PL area extending over several $\mu$m around the generation spot. This PL was collected by the same objective and split into two beams with intensity proportional to its left ($I_L$) and right ($I_R$) circular components, which were then imaged by a cooled charge-coupled detector (CCD) placed at the output of a spectrometer. The input slit of the spectrometer was placed parallel to the $y\parallel[1\overline{1}0]$ surface direction of the sample, so that the images of $I_R$ and $I_L$ were collected at the CCD with energy (vertical scale) and spatial resolution along $y$ (horizontal scale). In addition to spatially resolved, time-integrated profiles, we have also recorded time-resolved profiles by using a gated CCD camera synchronized with the laser pulses.

$I_R$ and $I_L$ were used to determine the time and spatial dependence of the carrier density, $n=I_R+I_L$ (we suppose $n=n_e=n_h$ because the QW is non-doped), as well as  the out-of-plane spin density $s_z=I_R-I_L$. As the spin polarization of the photoexcited holes is lost within a time (few picoseconds\cite{PhysRevLett.67.3432,PhysRevLett.89.146601}) much shorter than the \eh recombination lifetime, the difference between $I_R$ and $I_L$ only gives information about the out-of-plane component of the electron spin vector within the \eh ensemble. The out-of-plane spin polarization, $\rho_z$, was obtained according to $\rho_z=s_z/n=(I_R-I_L)/(I_R+I_L)$.  We limit the use of $\rho_z$, however, to time-integrated measurements, where the number of counts in the CCD for both $I_R$ and $I_L$ is significantly above the noise signal everywhere along $y$. As this is not always true in time-resolved experiments, in this case we only discuss $s_z$ to avoid dealing with meaningless values of the division $s_z/n$. 

\section{Results}\label{Sec_results}

\subsection{Carrier dynamics}\label{subsec_carrier}

Figures~\ref{setup}(b) and (c) compare time-integrated images of $I_R$ and $I_L$ recorded with spectral (vertical axis) and spatial resolution (horizontal axis) for $V_b=-1.5$~V. The position of the 1~$\mu$m-wide laser spot used for excitation is $y=0$. The dashed rectangles indicate the PL emitted by the electron-heavy hole (\textit{e-hh}) transition. Figure~\ref{setup}(d) shows the corresponding intensity profiles of $I_R$ and $I_L$ for the \textit{e-hh} transition as a function of the distance to the laser spot, while the dotted line displays, for comparison, the intensity profile of the tightly focused laser beam. As expected, the \textit{e-hh} emission region extends over an area much larger than the diameter of the excitation spot due to the radial ambipolar transport of photogenerated electrons and holes. The shape of the measured PL profile deviates from the expected one, symmetric to $y=0$, due to terraces at the sample surface, which originate from the growth on a surface tilted by an angle $\delta\theta$ with respect to the [111] direction.\cite{PVS272}

%*******************************************************************************
% Figure 2
\begin{figure}
\includegraphics[width=\columnwidth]{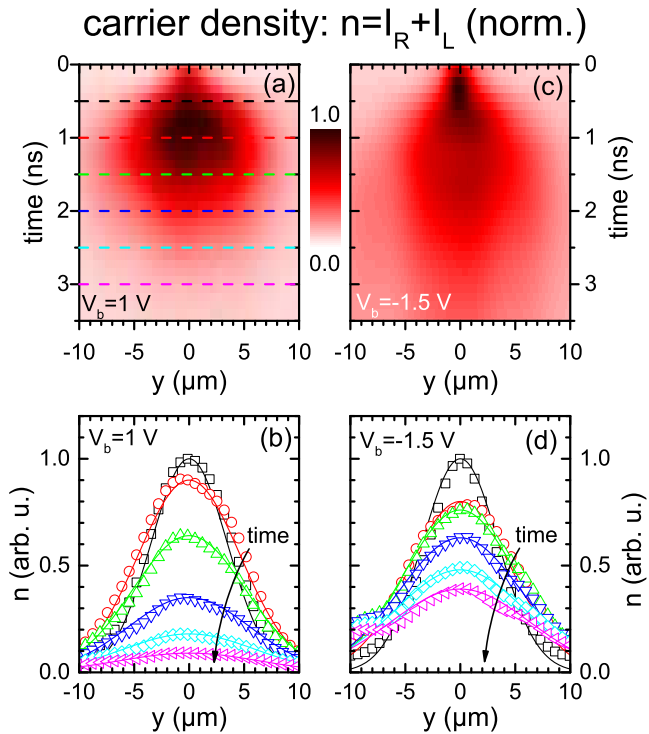}
\caption{Time and spatially resolved carrier density, $n$, along the $y$-direction after the laser pulse injects electron-hole pairs at $t=0$. Panel (a) shows the two-dimensional color profile for $V_b=1$~V, while panel (b) shows the spatial profiles of panel (a) for $t=0.5, 1.0, 1.5 ,2.0 ,2.5\mathrm{~and~}3.0$~ns (squares, circles, up triangles, down triangles, diamonds and left triangles). The solid lines are fittings to the data according to Eq.~\ref{Eq_2Ddiff}. Panels (c) and (d) show the corresponding results for $V_b=-1.5$~V. The experiment was performed at 30~K and a laser power of 20~$\mu$W. The curves are normalized to the maximum value at each panel.\label{charge_dynamics}}
\end{figure}
%*******************************************************************************

We have also performed time-resolved $\mu$-PL experiments using a time-gated CCD camera. Figure~\ref{charge_dynamics}(a) shows a two-dimensional color plot of the carrier density, $n$, as a function of the position along $y\parallel[1\overline{1}0]$ (horizontal scale), and the time delay with respect to the laser pulse (vertical scale) for $V_b=1$~V. Panel (b) displays the corresponding spatial profiles at the times marked with dashed lines in panel (a). After a fast initial expansion during the first nanosecond, the carrier motion slowers for longer time delays while $n$ also decays due to  \eh recombination. In contrast, when $V_b=-1.5$~V (corresponding to an electric field close to $E_z^{c}$), the carrier expansion lasts longer, cf. panels (c-d). We attribute this behavior to the reduction of the \eh recombination rate due to the electrically induced spatial separation of the electron and hole wave functions toward opposite barriers of the QW.

%*******************************************************************************
% Figure 3
\begin{figure}
\includegraphics[width=0.9\columnwidth]{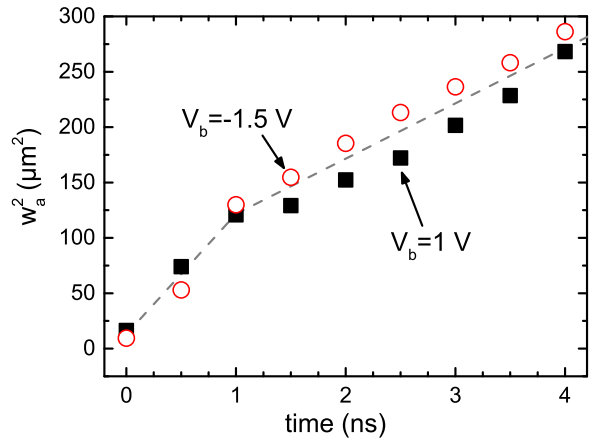}
\caption{Square of the full width at half maximum of $n$, $w_a$, as a function of time, obtained from the fitting of the curves of Fig.~\ref{charge_dynamics} to Eq.~\ref{Eq_2Ddiff}. The dashed lines are a guide to the eye.\label{charge_wa2vstime}}
\end{figure}
%*******************************************************************************

We can obtain the ambipolar carrier diffusion coefficient, $D_a$, and the \eh recombination time, $\tau_r$, from these time-resolved measurements by assuming that the optically generated carrier density, $n$, has an initial Gaussian shape. Under these conditions, the expansion of $n$ as a function of time, $t$, and radial distance, $r=|y|$, is described by the solution to a two-dimensional drift-diffusion model:\cite{Zhao_PRB79_115321_09}

\begin{equation}\label{Eq_2Ddiff}
n(r,t)=n_0\frac{w_0^{2}}{w_a^{2}(t)}\exp{\left(-\frac{4\ln(2)r^{2}}{w_a^{2}(t)}\right)}\exp{\left(-t/\tau_r\right)},
\end{equation}

\noindent where $n_0$ is the carrier density at $r=0$ and $t=0$, $w_0$ is the full width at half maximum (FWHM) of the initial carrier distribution, and $w_a(t)=[w_0^{2}+16\ln(2)D_at]^{1/2}$ is the FWHM of the profile as a function of time. Figure~\ref{charge_wa2vstime} shows values of $w_a^2$ obtained from the fits of Eq.~\ref{Eq_2Ddiff} to the profiles in Figure~\ref{charge_dynamics} at several time delays. The high quality of the fits is demonstrated by the solid lines superimposed on the experimental points in Figs.~\ref{charge_dynamics}(b) and \ref{charge_dynamics}(d). For both applied biases, $w_a^{2}(t)$ initially increases with a fast slope during the first nanosecond, followed by a slower evolution for larger time delays. We attribute the fast expansion just after the pulse ($t\leq1$~ns) to repulsive drift forces between carriers at the high density regions close to the generation spot. As the \eh pairs were non-resonantly generated by the laser, an initial higher carrier temperature with respect to the lattice also contributes to the fast expansion with an initially higher diffusion coefficient. The evolution at $t>1$~ns is due to the radial transport of the thermalized electrons and holes driven by diffusion. From the linear fitting of $w_a^2(t)$ in this regime, we obtain $D_a\approx45$~cm$^{2}$/s. It is remarkable that, although the increase of $E_z$ enhances the \eh recombination time from $\tau_r=2$~ns to 4~ns, $D_a$ is not significantly affected. To estimate the carrier scattering time, $\tau_p $, from $D_a$, we take into account that the Fermi energy for the carrier density present in the QW at $t>1$~ns is lower than the thermal energy. We can therefore apply the Einstein equation $D_a=\mu_ak_BT/q$, where $q$ is the electron charge, $k_B$ is the Boltzmann constant, and $\mu_a$ is the ambipolar mobility $\mu_a=2\mu_e\mu_{hh}/(\mu_e+\mu_{hh})$, whose value is intermediate between the electron and heavy hole mobilities, $\mu_e$ and $\mu_{hh}$ respectively. Our result is $\tau_p\approx3$~ps, in agreement with previously estimated values in a similar QW based on measurements of Dyakonov-Perel' spin dephasing time under homogeneous carrier concentration.\cite{PVS261,PVS272}

%*******************************************************************************
% Figure 4
\begin{figure}
\includegraphics[width=0.9\columnwidth]{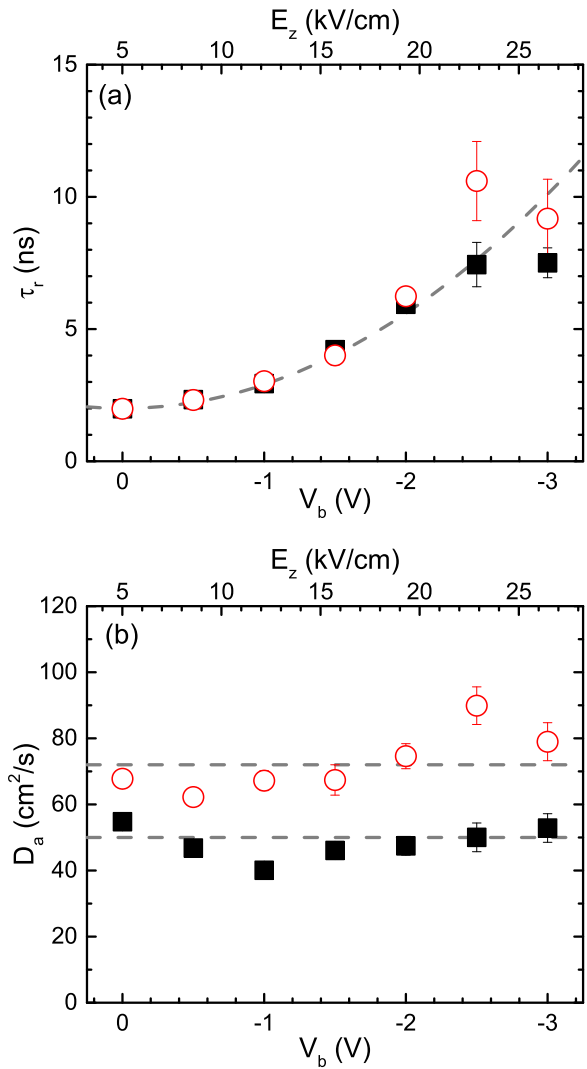}
\caption{Results of the fitting parameters of Eq.~\ref{Eq_2Ddiff} as a function of the applied bias for laser powers 40~$\mu$W (squares) and 80~$\mu$W (circles), measured at 40~K. (a) Carrier recombination time, $\tau_r$. (b) Ambipolar diffusion coefficient, $D_a$. The dashed lines in both panels are a guide to the eye.\label{charge_summary}}
\end{figure}
%*******************************************************************************

We have repeated the previous experiment for several voltage biases and laser powers, $P_{laser}$, to extract the dependence of $\tau_r$ and $D_a$ on $E_z$ and carrier density. Figure~\ref{charge_summary} shows (a) $\tau_r$ and (b) $D_a$ as a function of $V_b$ obtained from measurements performed at $P_{laser}=40~\mu$W (squares) and 80~$\mu$W (circles). The top horizontal scales depict the $E_z$ that corresponds to the applied $V_b$, which was determined from the energy shift of the PL due to the Quantum Confined Stark Effect (QCSE).\cite{PVS272} As expected, the spatial separation of electron and hole towards opposite QW barriers induced by $E_z$ increases the carrier lifetime. $D_a$, on the contrary, does not depend on applied bias. It increases, however, with the laser power, from 50~cm$^2$/s at 40~$\mu$W to 70~cm$^2$/s at 80~$\mu$W. We attribute the increase to the fact that a larger carrier concentration screens more efficiently the scattering centers that limit the carrier mobility in the structure.\cite{Remeika_PRL102_186803_09} An additional cause of this enhancement is that, for $P_{laser}\geq 80~\mu$W, the Fermi level approaches the thermal energy at the time range studied. Above this limit, $D_a$ becomes proportional to the carrier density. 

\subsection{Spin dynamics}\label{subsec_spin}

%*******************************************************************************
% Figure 5
\begin{figure}
\includegraphics[width=0.95\columnwidth]{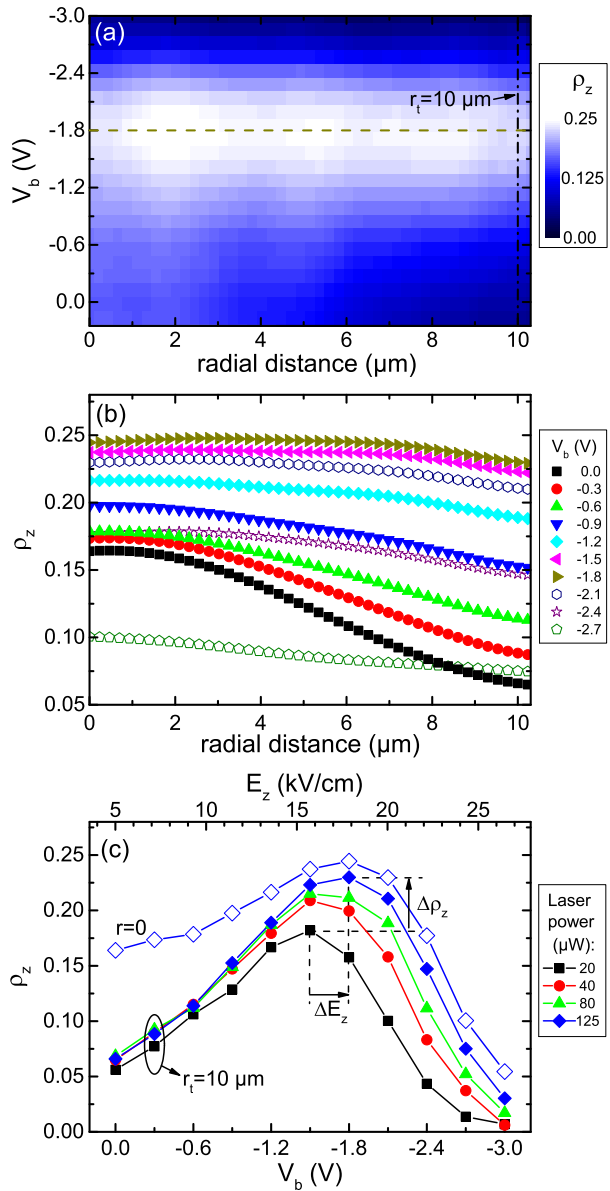}
\caption{(a) Time-integrated, out-of-plane spin polarization, $\rho_z$, as a function of radial distance from the generation spot, $r$, under different voltage biases, $V_b$. Measured at 30~K and a laser power 125~$\mu$W. (b) Radial profiles of $\rho_z$ for selected bias voltages of panel (a). (c) $\rho_z$ at $r=0$ (open symbols) and at $r_t=10$~$\mu$m (solid symbols) as a function of $V_b$ and its respective electric field, $E_z$. The different simbols correspond to laser powers of 20~$\mu$W (squares), 40~$\mu$W (circles), 80~$\mu$W (triangles) and 125~$\mu$W (diamonds). The lines are a guide to the eye.\label{spin_pol}}
\end{figure}
%*******************************************************************************

The large extension of the time-integrated PL profiles of Fig.~\ref{setup} relative to the laser spot diameter allows to determine the out-of-plane spin polarization, $\rho_z$, as the carriers move away from the excitation point. Figure~\ref{spin_pol}(a-b) shows $\rho_z$ vs. the radial distance, $r$, calculated from the time-integrated $I_R$ and $I_L$ profiles under different biases $V_b$. The experiment was performed using a laser power $P_{laser}=125~\mu$W. Figure~\ref{spin_pol}(c) compares the values of $\rho_z$ at the generation spot, $r=0$ (open diamonds), and at a radial transport distance, $r_t$, of 10~$\mu$m (solid diamonds) as a function of $V_b$ and its respective $E_z$. At $V_b=0$, $\rho_z$ decays from 0.16 at $r=0$ to 0.07 at $r_t$ due to spin precession around the SO-field between scattering events as the carriers move away from the generation spot. When the bias approaches $V_b=-1.8$~V, the linear Dresselhaus and the Rashba term cancel each other and the SO-interaction is suppressed. As a consequence, the electron ensemble expands without losing its initial out-of-plane spin polarization. Therefore, $\rho_z$ remains constant around $0.24$ all along the transport distance studied. At biases $V_b<-1.8$~V, the Rashba term overcompensates the Dresselhaus one. The SO-interaction becomes then active again and $\rho_z$ decreases. The observation of a maximum in $\rho_z$ at $V_b=-1.8$~V unambiguously establishes the BIA/SIA compensation as the mechanism for electrically induced spin transport enhancement in (111) QWs. The corresponding value of $E_z^{c}\approx~18$~kV/cm agrees well with PL results recorded in a similar sample using a larger illumination area.\cite{PVS261,PVS272} 

It is noteworthy that the $\rho_z$ profile for $r=0$ in Fig.~\ref{spin_pol}(c) is asymmetric with respect to the compensation bias $V_b=-1.8$~V. The asymmetry arises from the fact that, in a time-integrated PL measurement, $\rho_z$ depends on the ratio between the carrier recombination time, $\tau_r$, and the spin dephasing time, $\tau_s$, according to:

\begin{equation}\label{Eq_spinpol}
\rho_z=\rho_0/(1+\tau_r/\tau_s).
\end{equation}

\noindent Here, $\rho_0=0.25$ is the maximum electron spin polarization that can be achieved for excitation energies above the electron-light hole transition.\cite{Meier_84} While $\tau_r$ increases monotonously with $E_z$, cf. Fig.~\ref{charge_summary}(a), $\tau_s$ has a maximum at $E_z^c$ and decreases as the electric field moves away from the compensation field.\cite{Cartoixa05a, Vurgaftman_JAP97_53707_05, PVS261} Therefore, when $E_z$ exceeds $E_z^c$, $\tau_s<\tau_r$ and $\rho_z$ at $r=0$ goes to zero because the out-of-plane spin polarization is lost much before the electrons and holes recombine.

We have repeated the experiment of Fig.~\ref{spin_pol}(a) using several laser excitation powers, cf. Fig.~\ref{spin_pol}(c). As the number of optically injected carriers increases, we observe an enhancement of $\rho_z(E_z^c)$ at $r_t$, $\Delta\rho_z$. To understand this, we take into account that the photons detected at this point are emitted by \eh pairs that recombine just after traveling from the generation spot to $r_t$ in an average time $r_t^2/D_a$. We have shown in Fig.~\ref{charge_summary}(b) that $D_a$ increases with laser power, thus reducing the traveling time of the carriers. As a consequence, a larger fraction of electrons reaches $r_t$ before losing their out-of-plane spin polarization.

In addition, larger laser powers shift $E_z^c$ by an amount $\Delta E_z$ towards larger values. We exclude thermal effects as the cause of this shift: higher temperatures would allow the occupation of energy states with larger $k$-vectors, whose contribution to the cubic terms of $\bOmegaSO$ reduces $E_z^c$.\cite{Vurgaftman_JAP97_53707_05, PVS261, PVS272, Balocchi_NJ15_95016_13} We attribute the enhancement of $E^c_z$ to a partial screening of the electric field generated by the top gate. This screening comes from the accumulation of photoexcited electrons and holes at opposite walls of the QW as a response to the gate field.

%*******************************************************************************
% Figure 6
\begin{figure}
\includegraphics[width=\columnwidth]{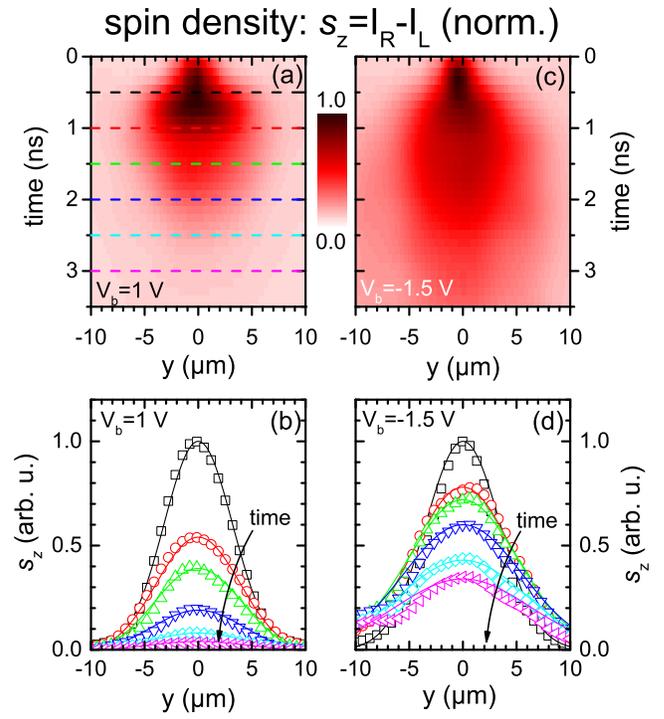}
\caption{Time and spatially resolved out-of-plane spin density, $s_z$, along the $y$-direction after the laser pulse injects electron-hole pairs at $t=0$. Panel (a) shows the two-dimensional color profiles for $V_b=1$~V, while panel (b) shows the spatial profiles for $t=0.5, 1.0, 1.5 ,2.0 ,2.5 ,3.0$~ns (squares, circles, up triangles, down triangles, diamonds and left triangles). The solid lines are simulations according to Eq.~\ref{Eq_szSimulation}. Panels (c-d) show the same as (a-b) for $V_b=-1.5$. The experiment was performed at 30~K and a laser power of 20~$\mu$W. The curves are normalized to the maximum value at each panel.
\label{spin_dynamics}}
\end{figure}
%*******************************************************************************

%*******************************************************************************
% Figure 7
\begin{figure}
\includegraphics[width=0.9\columnwidth]{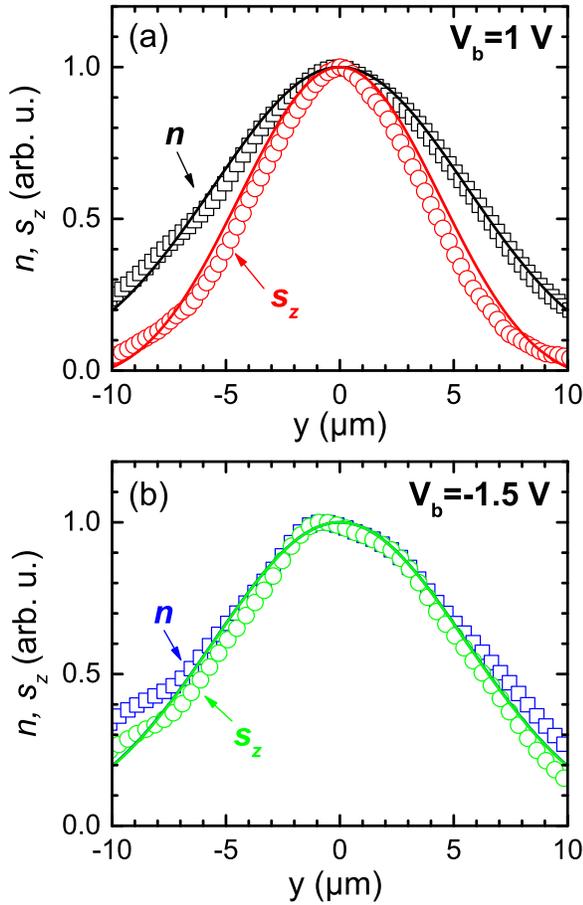}
\caption{Comparison of spatial profiles of $n$ (squares) and $s_z$ (circles) from Figs.~\ref{charge_dynamics} and \ref{spin_dynamics} for (a) $V_b=1$~V, and (b) $V_b=-1.5$~V, at a time delay of 2~ns. All curves are normalized to their maximum. The solid lines are simulations according to Eqs.~\ref{Eq_2Ddiff}, \ref{Eq_szSimulation} and \ref{Eq_angle}.
\label{spin_comparison}}
\end{figure}
%*******************************************************************************

%*******************************************************************************
% Figure 8
\begin{figure}
\includegraphics[width=0.9\columnwidth]{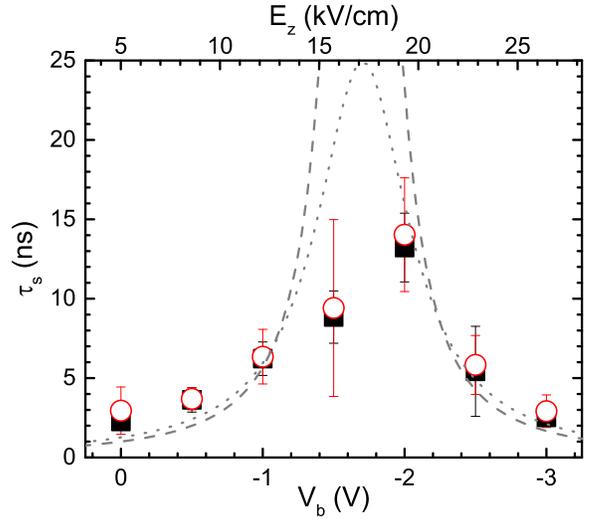}
\caption{Spin decoherence time, $\tau_s$, estimated from the fitting of $s_z$ by Eq.~\ref{Eq_szSimulation}, as a function of the applied bias $V_b$ for laser powers 40~$\mu$W (squares) and 80~$\mu$W (circles), measured at 40~K. The curves show, for comparison, the spin decoherence time in an homogeneous electron distribution for out-of-plane spins (dashed lines) and precessing spins (dotted line), estimated according to Ref.~\onlinecite{PVS261}.\label{spin_summary}}
\end{figure}
%*******************************************************************************

Finally, we have also measured the time-resolved radial expansion of the out-of-plane spin density, $s_z$, in the same way as we did for $n$ in subsection \ref{subsec_carrier}. Figure~\ref{spin_dynamics} shows the spatial- and time-resolved evolution of $s_z$ corresponding to the carrier dynamics of Figure~\ref{charge_dynamics}. As in the case of $n$, $s_z$ expands over an area much larger than the diameter of the excitation spot. A comparison of the temporal decay of $n$ and $s_z$ for $V_b=1$~V (panels (a-b) of Figures~\ref{charge_dynamics} and \ref{spin_dynamics}) shows that the amplitude of $s_z$ decreases slightly faster than $n$ due to the fact that the decay of $s_z$ includes spin dephasing in addition to carrier recombination. On the contrary, for the compensation bias $V_b=-1.5$~V (see panels (c-d) in the same figures), the amplitude decays of $n$ and $s_z$ are very similar. The latter is attributed to a negligible spin dephasing within the recombination lifetime of the carriers due to SO-compensation. The decay of the spin density must, therefore, follow the one of the carrier density. As the lifetime of the photogenerated carriers is only a few nanoseconds, the main mechanism limiting spin transport in intrinsic GaAs (111) QWs at SO-compensation is electron-hole recombination.

Interestingly, the spatial extension of $s_z$ also depends on $V_b$. Figure~\ref{spin_comparison} compares the normalized profiles of $n$ (squares) and $s_z$ (circles) at $t=2$~ns for (a) $V_b=1$~V and (b) $V_b=-1.5$~V. In the case of $V_b=1$~V, the profile of $s_z$, with a FWHM of $w_s=8.6~\mu$m, is clearly narrower than the one of $n$ ($w_n=12.7~\mu$m). When $V_b=-1.5$~V, in contrast, both profiles have a similar spatial extension of $w_n=13.7~\mu$m and $w_s=12.2~\mu$m, respectively. The discrepancy in the profile widths for biases away from SO-compensation is due to the coherent precession of the spin vectors around the SO-field during the two-dimensional expansion of the inhomogeneous electron population. When the Dresselhaus and Rashba terms fulfill special symmetries, the interplay between the Brownian motion of the carriers and the spin-orbit field can correlate the orientation of the spin vector at spatially separated points in the electron ensemble.\cite{Froltsov_PRB64_45311_01, Stanescu_PRB75_125307_07, Yang_PRB82_155324_10, Poshakinskiy_PRB92_45308_15} The most extreme example is the persistent spin helix observed in GaAs (001) QWs for a particular ratio between the Dresselhaus and Rashba terms.\cite{Bernevig_PRL97_236601_06, Koralek_N458_610_09, Walser2012} In GaAs (111) QWs, the linear Dresselhaus contribution and the Rashba one have the same symmetry. Therefore, the SO-interaction at low temperatures can be treated as a single in-plane term with the symmetry of the Rashba contribution. In this case, the two-dimensional expansion of an initially narrow out-of-plane spin ensemble leads to oscillations of $s_z$ along $k$ in the momentum space that are described by a Bessel function.\cite{Stanescu_PRB75_125307_07} These Bessel-like oscillations are also present in the analytic solution of $s_z(r,t)$ at times fulfilling $t\gg r^2/D$, with the first node appearing at $r_n=2.4837\sqrt{D\tau_s}$.\cite{Froltsov_PRB64_45311_01, Stanescu_PRB75_125307_07, Poshakinskiy_PRB92_45308_15, Altmann_PRB92_235304_15} In our experiments, the maximum radial distance is $r_t=10~\mu$m and the ambipolar diffusion coefficient is $D_a\approx50$~cm$^2$/s, thus leading to $r_t^2/D_a\approx20$~ns. The \eh recombination time in our sample is, however, only a few nanoseconds, cf. Fig.~\ref{charge_summary}(a). The spin density profiles of Figs.~\ref{spin_dynamics} and \ref{spin_comparison} correspond, therefore, to the opposite limit $t\ll r^2/D$, where no analytic solution in the spatial domain exists.\cite{Stanescu_PRB75_125307_07} In addition, the value of $\tau_s$ for the range of applied voltage biases (cf. Fig.~\ref{spin_summary}) implies that the first node appears at radial distances exceeding $r_t$. This means that the optically injected \eh pairs disappear well before the oscillation of $s_z$ along $r$ can be observed.

\section{Discussion}\label{Sec_discussion}

It is possible to obtain an analytic approximation to $s_z$ in the real space for $t\ll r^2/D$ if we take into account that, for the short carrier lifetime and weak SO-interaction of our experiment, the precession angle of the spin vectors in the ensemble is expected to be small. Under this assumption, the spatial profile of $s_z$ will resemble that of $n$ and can be approximated by:
 
\begin{equation}\label{Eq_szSimulation}
s_z(r,t)=n(r,t)\exp(-t/\tau_s)\cos[\theta(r,t)].
\end{equation}

Here, the cosine accounts for the projection along $z$ of the tilted spin vector due to its slight precession around $\bOmegaSO$ during the radial expansion.\footnote{Although a Bessel function is a more appropriate solution for a radially symmetric differential equation than a harmonic oscillation, both functions behave similar for $r\ll r_n$.} To determine the precession angle $\theta(r,t)$ of the spin ensemble, we proceed as follows: during an infinitesimal time interval $dt$, an electron spin vector moving with an in-plane momentum $\mathbf{k}$ rotates an infinitesimal angle $d\mathbf{\theta}=\bOmegaSO(\mathbf{k}) dt$. Taking into account only the linear in-plane $k$ terms in Eq.~\ref{Eq_BIA111}, and using the relations $\hbar\mathbf{k}=m^*\mathbf{v}$ and $\hbar^2\langle k^2\rangle\approx 2m^* k_BT$, we can express $\bOmegaSO$ as:

\begin{equation}\label{Eq_OmegaSOv}
\bOmegaSO(r,t)= g(T,E_z)(\mathbf{v}\times\unitary{z}),
\end{equation}

\begin{equation}
g(T,E_z)=\frac{2m^*\gamma}{\hbar^2\sqrt{3}}\left[\langle k_z^2\rangle -\frac{m^*}{2\hbar^2}k_BT-\sqrt{3}\frac{r_{41}}{\gamma}E_z\right].
\end{equation}

\noindent Here, $g(T,E_z)$ is the SO-precession angle per unit distance, and it fulfills $g(T,E_z^c)=0$. In our model, we assume that, during the early stage of the expansion process, the carriers generated within the laser spot follow a trajectory close to the radial one. This means that the instantaneous velocity of each carrier can be expressed as $\mathbf{v}=v(r,t)\hat{\mathbf{r}}+\mathbf{v}^*$. Here, $v(r,t)$ is the average radial velocity of the carrier ensemble, which is determined by the solution of the drift-diffusion differential equation, cf. Eq.~\ref{Eq_2Ddiff}. $\mathbf{v}^*$, on the contrary, is a perturbative random term that changes magnitude and direction after each scattering event. According to these definitions, the average precession angle, $\langle d\theta\rangle$, of the carriers contained in an infinitesimal area at $\mathbf{r}$ and $t$ during an infinitesimal time interval $dt$ is then just:

\begin{equation}
\langle\bOmegaSO\rangle dt=g(T,E_z)v(r,t)dt(\unitary{r}\times\unitary{z}),
\end{equation}

\noindent because $\langle\mathbf{v}^*\rangle=0$. Therefore, $v(r,t)$ determines the infinitesimal precession angle around which the spin ensemble coherently precesses at each position and time, while $\mathbf{v}^*$ is responsible for the spread in precession angles that gives rise to the spin dephasing time $\tau_s$.

To obtain $\theta(r,t)$, we just add $\langle d\theta\rangle$ along the radial trajectory of the carriers from their initial position within the gaussian distribution $r_0$ at $t=0$ until they reach $r$ at $t$. By doing this, we are neglecting the effect on $\theta(r,t)$ of carrier diffusion along other trajectories than the radial one. Taking into account that $v=dr/dt$, the integral is simply given by:

\begin{equation}
\theta(r,t)=g(T,E_z)\int_{r_0}^r dr'=g(T,E_z)(r-r_0).
\end{equation}

\noindent The problem reduces to finding the initial radial position, $r_0$, of particles reaching the position $r$ at the time $t$. To do this, we take into account that the radial current density $j=qnv$ in a drift-diffusion process depends on $r$ and $t$ according to:

\begin{equation}\label{Eq_Fickslaw}
qn(r,t)v(r,t)=-qD_a\frac{\partial n(r,t)}{\partial r}+j_{d}(r,t).
\end{equation}

\noindent Here, the first term in the right side of the equation is the diffusion current determined by Fick's first law. The second term, $j_d=\mu_aF_r$, represents the radial drift current induced by the radial repulsive forces, $F_r$, between excitons and free carriers at the high density regions close to the generation spot. As discussed in Sec.~\ref{subsec_carrier}, $j_d$ is only important during a short time after the laser pulse, where it is responsible of the fast carrier expansion observed for $t<1$~ns. It can be shown that, due to the shape and radial symmetry of $n$, the radial profile of $j_d$ is similar to that of $\partial n/\partial r$.\footnote{The repulsive forces at $r=0$ must be $F_r=0$ because of the radial symmetry of the carrier density. Due its gaussian shape, the carrier concentration is the largest at the center and decays radially towards zero. Therefore, $F_r$ must initially increase with $r$ until reaching a maximum around $w_n/2$, and then decay towards $F_r\rightarrow0$.} We can therefore take $j_d$ into account by supposing that it is $j_d\approx-qD_{d}\partial n/\partial r$ for $t<1$~ns and neglect it afterwards. The carriers then expand with an effective diffusion coefficient $D_{eff}=D_a+D_d>D_a$ for $t<1$~ns, while for $t>1$ the expansion is purely diffusive and determined only by $D_a$.

Introducing now the carrier density, $n$, of Eq.~\ref{Eq_2Ddiff} and its derivative $\partial n/\partial r$ into Eq.~\ref{Eq_Fickslaw} and integrating $v=dr/dt$, we obtain the following relation between $r_0$ and $r$:

\begin{equation}\label{Eq_r}
\frac{r_0}{w_0}=\frac{r}{w_a(t)},
\end{equation}

\noindent which leads to the final expression for $\theta(r,t)$:

\begin{equation}\label{Eq_angle}
\theta(r,t)=g(T,E_z)r\left[1-\frac{w_0}{w_a(t)}\right].
\end{equation}

\noindent The precession angle at each point, therefore, increases with time from zero towards a finite value that saturates at long times. Remarkably, this result is qualitatively similar to the obtained for an initial spin polarization with finite spatial extension close to the spin helix regime,\cite{Walser2012} but differs in the fact that the width enters as $1-(w_0/w_a)^2$ in the spin helix case.\cite{Salis_PRB89_45304_14} We attribute the origin of this discrepancy to the different symmetry of the SO-interaction studied in each case.

To simulate $g(T,E_z)$, we have assumed that $\gamma=17$~eV\AA$^3$, $r_{41}=6$~e\AA$^2$ and $d_\mathrm{eff}=28.6$~nm ($d_\mathrm{eff}$ takes into account both the QW thickness and the penetration length into the barriers) from our previous calculations of Ref.~\onlinecite{PVS261}. The initial carrier density profile, $n(r,0)$, is a 2D gaussian function with $w_0=3.6~\mu$m, in agreement with the measured FWHM of the carrier density at $t=0$. $n$ and $s_z$ evolve according to Eqs.~\ref{Eq_2Ddiff}, \ref{Eq_szSimulation} and \ref{Eq_angle}, with $D_{eff}=98$~cm$^2$/s for $t<1$~ns and $D_a=45$~cm$^3$/s for $t>1$~ns, as estimated from the slope of $w_a^2$ in Fig.~\ref{charge_wa2vstime} at each time interval. The solid lines in Figs.~\ref{charge_dynamics}, \ref{spin_dynamics} and \ref{spin_comparison} display the calculated spatial profiles for $n$ and $s_z$ at the corresponding time delays, showing a reasonable agreement with the experiment. For $E_z$ away from the SO-compensation, the reduced width of $s_z$ with respect to $n$ is due to the coherent precession of the spin vector around $\bOmegaSO$. When $E_z\approx E_z^c$, then $g(T,E_z^c)=0$, the spin precession is suppressed, and therefore the FWHM of $s_z$ and $n$ become similar. 

In the calculation, we have taken into account neither the cubic terms in $k$ of $\bOmegaD$, nor correction terms related to the slightly tilt of the substrate surface away from the [111] direction. These additional contributions introduce out-of-plane SO-field components and, most important, they break the symmetry between $\bOmegaD$ and $\bOmegaR$. As a consequence, the in-plane $\OmegaSO$ does not cancel simultaneously for all the electrons in the ensemble, but only for a certain fraction, thus avoiding the extension of the measured $s_z$ profile to fully equal that of $n$ at $E_z^c$, cf. Fig.~\ref{spin_comparison}(b).

The fits of the temporal decay of $s_z$ by Eq.~\ref{Eq_szSimulation} yields the dependence of $\tau_s$ on $E_z$ and on the laser power shown in Fig.~\ref{spin_summary}. The dashed and dotted lines show, for comparison, the calculated DP dephasing lifetime for out-of-plane and precessing spins in an homogeneous spin ensemble.\cite{PVS261} As expected, $\tau_s$ has a maximum at $E_z^c\approx18$~kV/cm, where it fulfills $\tau_s\gg\tau_r$, and decreases as the electric field moves away from SO-compensation. Finally, we would like to remark that, for $E_z>20$~kV/cm, $\tau_s$ lies below $\tau_r$, as expected from the behavior of the time-integrated $\rho_z$ discussed in Fig.~\ref{spin_pol}.

\section{Summary}\label{Sec_summary}

We have shown that the application of a transverse electric field allows the transport of out-of-plane spin polarized electrons in an intrinsic GaAs (111) QW over distances exceeding 10~$\mu$m. We attribute the long transport distance to the compensation of the in-plane Dresselhaus SO-interaction by the Rashba one generated by the electric field, which cancels the precession of the spin vector around the SO-field during the radial expansion of the photogenerated \eh ensemble. At SO-compensation, the main mechanism limiting the transport distance of the spin-polarized electrons is their recombination with holes, which takes place in typically a few nanoseconds. The favorable properties for spin transport of GaAs (111) QWs make them an excellent platform for further studies of the spin dynamics, as well as for spintronics applications.

\section*{Acknowledgements}
We thank Dr. Ramsteiner for useful discussions as well as M. H\"oricke and S. Rauwerdink for MBE growth and sample processing. We gratefully acknowledge financial support from the German DFG (priority program  SSP1285).

%\bibliography{literature,mypapers}

\begin{thebibliography}{54}%
\makeatletter
\providecommand \@ifxundefined [1]{%
 \@ifx{#1\undefined}
}%
\providecommand \@ifnum [1]{%
 \ifnum #1\expandafter \@firstoftwo
 \else \expandafter \@secondoftwo
 \fi
}%
\providecommand \@ifx [1]{%
 \ifx #1\expandafter \@firstoftwo
 \else \expandafter \@secondoftwo
 \fi
}%
\providecommand \natexlab [1]{#1}%
\providecommand \enquote  [1]{``#1''}%
\providecommand \bibnamefont  [1]{#1}%
\providecommand \bibfnamefont [1]{#1}%
\providecommand \citenamefont [1]{#1}%
\providecommand \href@noop [0]{\@secondoftwo}%
\providecommand \href [0]{\begingroup \@sanitize@url \@href}%
\providecommand \@href[1]{\@@startlink{#1}\@@href}%
\providecommand \@@href[1]{\endgroup#1\@@endlink}%
\providecommand \@sanitize@url [0]{\catcode `\\12\catcode `\$12\catcode
  `\&12\catcode `\#12\catcode `\^12\catcode `\_12\catcode `\%12\relax}%
\providecommand \@@startlink[1]{}%
\providecommand \@@endlink[0]{}%
\providecommand \url  [0]{\begingroup\@sanitize@url \@url }%
\providecommand \@url [1]{\endgroup\@href {#1}{\urlprefix }}%
\providecommand \urlprefix  [0]{URL }%
\providecommand \Eprint [0]{\href }%
\providecommand \doibase [0]{http://dx.doi.org/}%
\providecommand \selectlanguage [0]{\@gobble}%
\providecommand \bibinfo  [0]{\@secondoftwo}%
\providecommand \bibfield  [0]{\@secondoftwo}%
\providecommand \translation [1]{[#1]}%
\providecommand \BibitemOpen [0]{}%
\providecommand \bibitemStop [0]{}%
\providecommand \bibitemNoStop [0]{.\EOS\space}%
\providecommand \EOS [0]{\spacefactor3000\relax}%
\providecommand \BibitemShut  [1]{\csname bibitem#1\endcsname}%
\let\auto@bib@innerbib\@empty
%</preamble>
\bibitem [{\citenamefont {Awschalom}\ \emph {et~al.}(2002)\citenamefont
  {Awschalom}, \citenamefont {Loss},\ and\ \citenamefont
  {Samarth}}]{Awschalom2002}%
  \BibitemOpen
  \bibfield  {author} {\bibinfo {author} {\bibfnamefont {D.}~\bibnamefont
  {Awschalom}}, \bibinfo {author} {\bibfnamefont {D.}~\bibnamefont {Loss}}, \
  and\ \bibinfo {author} {\bibfnamefont {N.}~\bibnamefont {Samarth}},\
  }\href@noop {} {\emph {\bibinfo {title} {Semiconductor spintronics and
  quantum computation}}},\ Nanoscience and technology.\ (\bibinfo  {publisher}
  {Springer},\ \bibinfo {address} {Berlin ; New York},\ \bibinfo {year}
  {2002})\BibitemShut {NoStop}%
\bibitem [{\citenamefont {Zutic}\ \emph {et~al.}(2004)\citenamefont {Zutic},
  \citenamefont {Fabian},\ and\ \citenamefont {Sarma}}]{Zutic_ROMP76_323_04}%
  \BibitemOpen
  \bibfield  {author} {\bibinfo {author} {\bibfnamefont {I.}~\bibnamefont
  {Zutic}}, \bibinfo {author} {\bibfnamefont {J.}~\bibnamefont {Fabian}}, \
  and\ \bibinfo {author} {\bibfnamefont {S.~D.}\ \bibnamefont {Sarma}},\
  }\href@noop {} {\bibfield  {journal} {\bibinfo  {journal} {Rev. Mod. Phys.}\
  }\textbf {\bibinfo {volume} {76}},\ \bibinfo {pages} {323} (\bibinfo {year}
  {2004})}\BibitemShut {NoStop}%
\bibitem [{\citenamefont {D'yakonov}\ and\ \citenamefont
  {Perel'}(1972)}]{Dyakonov_SPSS13_3023_72}%
  \BibitemOpen
  \bibfield  {author} {\bibinfo {author} {\bibfnamefont {M.~I.}\ \bibnamefont
  {D'yakonov}}\ and\ \bibinfo {author} {\bibfnamefont {V.~I.}\ \bibnamefont
  {Perel'}},\ }\href@noop {} {\bibfield  {journal} {\bibinfo  {journal} {Sov.
  Phys. Solid State}\ }\textbf {\bibinfo {volume} {13}},\ \bibinfo {pages}
  {3023} (\bibinfo {year} {1972})}\BibitemShut {NoStop}%
\bibitem [{\citenamefont {D'yakonov}\ and\ \citenamefont
  {Kachorovskii}(1986)}]{Dyakonov_SPS20_110_86}%
  \BibitemOpen
  \bibfield  {author} {\bibinfo {author} {\bibfnamefont {M.~I.}\ \bibnamefont
  {D'yakonov}}\ and\ \bibinfo {author} {\bibfnamefont {V.~Y.~Y.}\ \bibnamefont
  {Kachorovskii}},\ }\href@noop {} {\bibfield  {journal} {\bibinfo  {journal}
  {Sov. Phys. Semicond.}\ }\textbf {\bibinfo {volume} {20}},\ \bibinfo {pages}
  {110} (\bibinfo {year} {1986})}\BibitemShut {NoStop}%
\bibitem [{\citenamefont {Dresselhaus}(1955)}]{Dresselhaus55a}%
  \BibitemOpen
  \bibfield  {author} {\bibinfo {author} {\bibfnamefont {G.}~\bibnamefont
  {Dresselhaus}},\ }\href {\doibase 10.1103/PhysRev.100.580} {\bibfield
  {journal} {\bibinfo  {journal} {Phys. Rev.}\ }\textbf {\bibinfo {volume}
  {100}},\ \bibinfo {pages} {580} (\bibinfo {year} {1955})}\BibitemShut
  {NoStop}%
\bibitem [{\citenamefont {Christensen}\ and\ \citenamefont
  {Cardona}(1984)}]{Christensen_SSC51_491_84}%
  \BibitemOpen
  \bibfield  {author} {\bibinfo {author} {\bibfnamefont {N.}~\bibnamefont
  {Christensen}}\ and\ \bibinfo {author} {\bibfnamefont {M.}~\bibnamefont
  {Cardona}},\ }\href {\doibase http://dx.doi.org/10.1016/0038-1098(84)91019-6}
  {\bibfield  {journal} {\bibinfo  {journal} {Solid State Communications}\
  }\textbf {\bibinfo {volume} {51}},\ \bibinfo {pages} {491 } (\bibinfo {year}
  {1984})}\BibitemShut {NoStop}%
\bibitem [{\citenamefont {Cardona}\ \emph {et~al.}(1988)\citenamefont
  {Cardona}, \citenamefont {Christensen},\ and\ \citenamefont
  {Fasol}}]{Cardona_PRB38_1806_88}%
  \BibitemOpen
  \bibfield  {author} {\bibinfo {author} {\bibfnamefont {M.}~\bibnamefont
  {Cardona}}, \bibinfo {author} {\bibfnamefont {N.~E.}\ \bibnamefont
  {Christensen}}, \ and\ \bibinfo {author} {\bibfnamefont {G.}~\bibnamefont
  {Fasol}},\ }\href@noop {} {\bibfield  {journal} {\bibinfo  {journal} {Phys.
  Rev. B}\ }\textbf {\bibinfo {volume} {38}},\ \bibinfo {pages} {1806}
  (\bibinfo {year} {1988})}\BibitemShut {NoStop}%
\bibitem [{\citenamefont {Eppenga}\ and\ \citenamefont
  {Schuurmans}(1988)}]{Eppenga_PRB37_10923_88}%
  \BibitemOpen
  \bibfield  {author} {\bibinfo {author} {\bibfnamefont {R.}~\bibnamefont
  {Eppenga}}\ and\ \bibinfo {author} {\bibfnamefont {M.~F.~H.}\ \bibnamefont
  {Schuurmans}},\ }\href@noop {} {\bibfield  {journal} {\bibinfo  {journal}
  {Phys. Rev. B}\ }\textbf {\bibinfo {volume} {37}},\ \bibinfo {pages} {10923}
  (\bibinfo {year} {1988})}\BibitemShut {NoStop}%
\bibitem [{\citenamefont {Bychkov}\ and\ \citenamefont
  {Rashba}(1984)}]{Bychkov1984}%
  \BibitemOpen
  \bibfield  {author} {\bibinfo {author} {\bibfnamefont {Y.~A.}\ \bibnamefont
  {Bychkov}}\ and\ \bibinfo {author} {\bibfnamefont {E.~I.}\ \bibnamefont
  {Rashba}},\ }\href {\doibase doi:10.1088/0022-3719/17/33/015} {\bibfield
  {journal} {\bibinfo  {journal} {J. Phys. C}\ }\textbf {\bibinfo {volume}
  {17}},\ \bibinfo {pages} {6039} (\bibinfo {year} {1984})}\BibitemShut
  {NoStop}%
\bibitem [{\citenamefont {Winkler}(2003)}]{Winkler03a}%
  \BibitemOpen
  \bibfield  {author} {\bibinfo {author} {\bibfnamefont {R.}~\bibnamefont
  {Winkler}},\ }\href@noop {} {\emph {\bibinfo {title} {Spin-Orbit Coupling
  Effects in Two-Dimensional Electron and Hole Systems}}},\ Vol.\ \bibinfo
  {volume} {191}\ (\bibinfo  {publisher} {Springer},\ \bibinfo {address}
  {Berlin},\ \bibinfo {year} {2003})\BibitemShut {NoStop}%
\bibitem [{\citenamefont {Cartoixa}\ \emph {et~al.}(2005)\citenamefont
  {Cartoixa}, \citenamefont {Ting},\ and\ \citenamefont {Chang}}]{Cartoixa05a}%
  \BibitemOpen
  \bibfield  {author} {\bibinfo {author} {\bibfnamefont {X.}~\bibnamefont
  {Cartoixa}}, \bibinfo {author} {\bibfnamefont {D.~Z.-Y.}\ \bibnamefont
  {Ting}}, \ and\ \bibinfo {author} {\bibfnamefont {Y.-C.}\ \bibnamefont
  {Chang}},\ }\href@noop {} {\bibfield  {journal} {\bibinfo  {journal} {Phys.
  Rev. B}\ }\textbf {\bibinfo {volume} {71}},\ \bibinfo {pages} {045313}
  (\bibinfo {year} {2005})}\BibitemShut {NoStop}%
\bibitem [{\citenamefont {Vurgaftman}\ and\ \citenamefont
  {Meyer}(2005)}]{Vurgaftman_JAP97_53707_05}%
  \BibitemOpen
  \bibfield  {author} {\bibinfo {author} {\bibfnamefont {I.}~\bibnamefont
  {Vurgaftman}}\ and\ \bibinfo {author} {\bibfnamefont {J.~R.}\ \bibnamefont
  {Meyer}},\ }\href {\doibase 10.1063/1.1858876} {\bibfield  {journal}
  {\bibinfo  {journal} {J. Appl. Phys.}\ }\textbf {\bibinfo {volume} {97}},\
  \bibinfo {pages} {053707} (\bibinfo {year} {2005})}\BibitemShut {NoStop}%
\bibitem [{\citenamefont {Balocchi}\ \emph {et~al.}(2011)\citenamefont
  {Balocchi}, \citenamefont {Duong}, \citenamefont {Renucci}, \citenamefont
  {Liu}, \citenamefont {Fontaine}, \citenamefont {Amand}, \citenamefont
  {Lagarde},\ and\ \citenamefont {Marie}}]{Balocchi2011}%
  \BibitemOpen
  \bibfield  {author} {\bibinfo {author} {\bibfnamefont {A.}~\bibnamefont
  {Balocchi}}, \bibinfo {author} {\bibfnamefont {Q.~H.}\ \bibnamefont {Duong}},
  \bibinfo {author} {\bibfnamefont {P.}~\bibnamefont {Renucci}}, \bibinfo
  {author} {\bibfnamefont {B.~L.}\ \bibnamefont {Liu}}, \bibinfo {author}
  {\bibfnamefont {C.}~\bibnamefont {Fontaine}}, \bibinfo {author}
  {\bibfnamefont {T.}~\bibnamefont {Amand}}, \bibinfo {author} {\bibfnamefont
  {D.}~\bibnamefont {Lagarde}}, \ and\ \bibinfo {author} {\bibfnamefont
  {X.}~\bibnamefont {Marie}},\ }\href@noop {} {\bibfield  {journal} {\bibinfo
  {journal} {Phys Rev. Lett.}\ }\textbf {\bibinfo {volume} {107}},\ \bibinfo
  {pages} {136604} (\bibinfo {year} {2011})}\BibitemShut {NoStop}%
\bibitem [{\citenamefont {Biermann}\ \emph {et~al.}(2012)\citenamefont
  {Biermann}, \citenamefont {Hern\'andez-M\'inguez}, \citenamefont {Hey},\ and\
  \citenamefont {Santos}}]{PVS257}%
  \BibitemOpen
  \bibfield  {author} {\bibinfo {author} {\bibfnamefont {K.}~\bibnamefont
  {Biermann}}, \bibinfo {author} {\bibfnamefont {A.}~\bibnamefont
  {Hern\'andez-M\'inguez}}, \bibinfo {author} {\bibfnamefont {R.}~\bibnamefont
  {Hey}}, \ and\ \bibinfo {author} {\bibfnamefont {P.~V.}\ \bibnamefont
  {Santos}},\ }\href@noop {} {\bibfield  {journal} {\bibinfo  {journal} {J.
  Appl. Phys.}\ }\textbf {\bibinfo {volume} {112}},\ \bibinfo {pages} {083913}
  (\bibinfo {year} {2012})}\BibitemShut {NoStop}%
\bibitem [{\citenamefont {Hern{\'a}ndez-M{\'i}nguez}\ \emph
  {et~al.}(2012)\citenamefont {Hern{\'a}ndez-M{\'i}nguez}, \citenamefont
  {Biermann}, \citenamefont {Hey},\ and\ \citenamefont {Santos}}]{PVS261}%
  \BibitemOpen
  \bibfield  {author} {\bibinfo {author} {\bibfnamefont {A.}~\bibnamefont
  {Hern{\'a}ndez-M{\'i}nguez}}, \bibinfo {author} {\bibfnamefont
  {K.}~\bibnamefont {Biermann}}, \bibinfo {author} {\bibfnamefont
  {R.}~\bibnamefont {Hey}}, \ and\ \bibinfo {author} {\bibfnamefont {P.~V.}\
  \bibnamefont {Santos}},\ }\href {\doibase 10.1103/PhysRevLett.109.266602}
  {\bibfield  {journal} {\bibinfo  {journal} {Phys. Rev. Lett.}\ }\textbf
  {\bibinfo {volume} {109}},\ \bibinfo {pages} {266602} (\bibinfo {year}
  {2012})}\BibitemShut {NoStop}%
\bibitem [{\citenamefont {Sun}\ \emph {et~al.}(2010)\citenamefont {Sun},
  \citenamefont {Zhang},\ and\ \citenamefont {Wu}}]{Sun_a__10}%
  \BibitemOpen
  \bibfield  {author} {\bibinfo {author} {\bibfnamefont {B.~Y.}\ \bibnamefont
  {Sun}}, \bibinfo {author} {\bibfnamefont {P.}~\bibnamefont {Zhang}}, \ and\
  \bibinfo {author} {\bibfnamefont {M.~W.}\ \bibnamefont {Wu}},\ }\href@noop {}
  {\bibfield  {journal} {\bibinfo  {journal} {J. Appl. Phys.}\ }\textbf
  {\bibinfo {volume} {108}},\ \bibinfo {pages} {093709} (\bibinfo {year}
  {2010})}\BibitemShut {NoStop}%
\bibitem [{\citenamefont {Wang}\ \emph
  {et~al.}(2013{\natexlab{a}})\citenamefont {Wang}, \citenamefont {Balocchi},
  \citenamefont {Lagarde}, \citenamefont {Zhu}, \citenamefont {Amand},
  \citenamefont {Renucci}, \citenamefont {Shi}, \citenamefont {Wang},
  \citenamefont {Liu},\ and\ \citenamefont {Marie}}]{Wang_APL102_242408_13}%
  \BibitemOpen
  \bibfield  {author} {\bibinfo {author} {\bibfnamefont {G.}~\bibnamefont
  {Wang}}, \bibinfo {author} {\bibfnamefont {A.}~\bibnamefont {Balocchi}},
  \bibinfo {author} {\bibfnamefont {D.}~\bibnamefont {Lagarde}}, \bibinfo
  {author} {\bibfnamefont {C.~R.}\ \bibnamefont {Zhu}}, \bibinfo {author}
  {\bibfnamefont {T.}~\bibnamefont {Amand}}, \bibinfo {author} {\bibfnamefont
  {P.}~\bibnamefont {Renucci}}, \bibinfo {author} {\bibfnamefont {Z.~W.}\
  \bibnamefont {Shi}}, \bibinfo {author} {\bibfnamefont {W.~X.}\ \bibnamefont
  {Wang}}, \bibinfo {author} {\bibfnamefont {B.~L.}\ \bibnamefont {Liu}}, \
  and\ \bibinfo {author} {\bibfnamefont {X.}~\bibnamefont {Marie}},\
  }\href@noop {} {\bibfield  {journal} {\bibinfo  {journal} {App. Phys. Lett.}\
  }\textbf {\bibinfo {volume} {102}},\ \bibinfo {pages} {242408} (\bibinfo
  {year} {2013}{\natexlab{a}})}\BibitemShut {NoStop}%
\bibitem [{\citenamefont {Balocchi}\ \emph {et~al.}(2013)\citenamefont
  {Balocchi}, \citenamefont {Amand}, \citenamefont {Wang}, \citenamefont {Liu},
  \citenamefont {Renucci}, \citenamefont {Duong},\ and\ \citenamefont
  {Marie}}]{Balocchi_NJ15_95016_13}%
  \BibitemOpen
  \bibfield  {author} {\bibinfo {author} {\bibfnamefont {A.}~\bibnamefont
  {Balocchi}}, \bibinfo {author} {\bibfnamefont {T.}~\bibnamefont {Amand}},
  \bibinfo {author} {\bibfnamefont {G.}~\bibnamefont {Wang}}, \bibinfo {author}
  {\bibfnamefont {B.~L.}\ \bibnamefont {Liu}}, \bibinfo {author} {\bibfnamefont
  {P.}~\bibnamefont {Renucci}}, \bibinfo {author} {\bibfnamefont {Q.~H.}\
  \bibnamefont {Duong}}, \ and\ \bibinfo {author} {\bibfnamefont
  {X.}~\bibnamefont {Marie}},\ }\href {\doibase 10.1088/1367-2630/15/9/095016}
  {\bibfield  {journal} {\bibinfo  {journal} {New. J. Phys.}\ }\textbf
  {\bibinfo {volume} {15}},\ \bibinfo {pages} {095016} (\bibinfo {year}
  {2013})}\BibitemShut {NoStop}%
\bibitem [{\citenamefont {Hern{\'a}ndez-M{\'i}nguez}\ \emph
  {et~al.}(2014)\citenamefont {Hern{\'a}ndez-M{\'i}nguez}, \citenamefont
  {Biermann}, \citenamefont {Hey},\ and\ \citenamefont {Santos}}]{PVS272}%
  \BibitemOpen
  \bibfield  {author} {\bibinfo {author} {\bibfnamefont {A.}~\bibnamefont
  {Hern{\'a}ndez-M{\'i}nguez}}, \bibinfo {author} {\bibfnamefont
  {K.}~\bibnamefont {Biermann}}, \bibinfo {author} {\bibfnamefont
  {R.}~\bibnamefont {Hey}}, \ and\ \bibinfo {author} {\bibfnamefont {P.~V.}\
  \bibnamefont {Santos}},\ }\href {\doibase 10.1002/pssb.201350202} {\bibfield
  {journal} {\bibinfo  {journal} {Phys. Status Solidi B}\ }\textbf {\bibinfo
  {volume} {251}},\ \bibinfo {pages} {1736} (\bibinfo {year}
  {2014})}\BibitemShut {NoStop}%
\bibitem [{\citenamefont {Kikkawa}\ and\ \citenamefont
  {Awschalom}(1999)}]{JKDA99a}%
  \BibitemOpen
  \bibfield  {author} {\bibinfo {author} {\bibfnamefont {J.~M.}\ \bibnamefont
  {Kikkawa}}\ and\ \bibinfo {author} {\bibfnamefont {D.~D.}\ \bibnamefont
  {Awschalom}},\ }\href@noop {} {\bibfield  {journal} {\bibinfo  {journal}
  {Nature}\ }\textbf {\bibinfo {volume} {397}},\ \bibinfo {pages} {139}
  (\bibinfo {year} {1999})}\BibitemShut {NoStop}%
\bibitem [{\citenamefont {Crooker}\ \emph {et~al.}(2005)\citenamefont
  {Crooker}, \citenamefont {M.~Furis}, \citenamefont {Adelmann}, \citenamefont
  {Smith}, \citenamefont {Palmstr{\"o}m},\ and\ \citenamefont
  {Crowell}}]{Crooker_Science_309_2191_05}%
  \BibitemOpen
  \bibfield  {author} {\bibinfo {author} {\bibfnamefont {S.~A.}\ \bibnamefont
  {Crooker}}, \bibinfo {author} {\bibfnamefont {.~X.~L.}\ \bibnamefont
  {M.~Furis}}, \bibinfo {author} {\bibfnamefont {C.}~\bibnamefont {Adelmann}},
  \bibinfo {author} {\bibfnamefont {D.~L.}\ \bibnamefont {Smith}}, \bibinfo
  {author} {\bibfnamefont {C.~J.}\ \bibnamefont {Palmstr{\"o}m}}, \ and\
  \bibinfo {author} {\bibfnamefont {P.~A.}\ \bibnamefont {Crowell}},\
  }\href@noop {} {\bibfield  {journal} {\bibinfo  {journal} {Science}\ }\textbf
  {\bibinfo {volume} {309}},\ \bibinfo {pages} {2191} (\bibinfo {year}
  {2005})}\BibitemShut {NoStop}%
\bibitem [{\citenamefont {Yu}\ \emph {et~al.}(2009)\citenamefont {Yu},
  \citenamefont {Zhang}, \citenamefont {Wang}, \citenamefont {Ni},
  \citenamefont {Niu},\ and\ \citenamefont {Lai}}]{Yu_APL94_202109_09}%
  \BibitemOpen
  \bibfield  {author} {\bibinfo {author} {\bibfnamefont {H.-L.}\ \bibnamefont
  {Yu}}, \bibinfo {author} {\bibfnamefont {X.-M.}\ \bibnamefont {Zhang}},
  \bibinfo {author} {\bibfnamefont {P.-F.}\ \bibnamefont {Wang}}, \bibinfo
  {author} {\bibfnamefont {H.-Q.}\ \bibnamefont {Ni}}, \bibinfo {author}
  {\bibfnamefont {Z.-C.}\ \bibnamefont {Niu}}, \ and\ \bibinfo {author}
  {\bibfnamefont {T.}~\bibnamefont {Lai}},\ }\href {\doibase 10.1063/1.3141483}
  {\bibfield  {journal} {\bibinfo  {journal} {App. Phys. Lett.}\ }\textbf
  {\bibinfo {volume} {94}},\ \bibinfo {pages} {202109} (\bibinfo {year}
  {2009})}\BibitemShut {NoStop}%
\bibitem [{\citenamefont {Quast}\ \emph {et~al.}(2009)\citenamefont {Quast},
  \citenamefont {Astakhov}, \citenamefont {Ossau}, \citenamefont {Molenkamp},
  \citenamefont {Heinrich}, \citenamefont {H\"ofling},\ and\ \citenamefont
  {Forchel}}]{Quast_PRB79_245207_09}%
  \BibitemOpen
  \bibfield  {author} {\bibinfo {author} {\bibfnamefont {J.-H.}\ \bibnamefont
  {Quast}}, \bibinfo {author} {\bibfnamefont {G.~V.}\ \bibnamefont {Astakhov}},
  \bibinfo {author} {\bibfnamefont {W.}~\bibnamefont {Ossau}}, \bibinfo
  {author} {\bibfnamefont {L.~W.}\ \bibnamefont {Molenkamp}}, \bibinfo {author}
  {\bibfnamefont {J.}~\bibnamefont {Heinrich}}, \bibinfo {author}
  {\bibfnamefont {S.}~\bibnamefont {H\"ofling}}, \ and\ \bibinfo {author}
  {\bibfnamefont {A.}~\bibnamefont {Forchel}},\ }\href {\doibase
  10.1103/PhysRevB.79.245207} {\bibfield  {journal} {\bibinfo  {journal} {Phys.
  Rev. B}\ }\textbf {\bibinfo {volume} {79}},\ \bibinfo {pages} {245207}
  (\bibinfo {year} {2009})}\BibitemShut {NoStop}%
\bibitem [{\citenamefont {Quast}\ \emph {et~al.}(2013)\citenamefont {Quast},
  \citenamefont {Henn}, \citenamefont {Kiessling}, \citenamefont {Ossau},
  \citenamefont {Molenkamp}, \citenamefont {Reuter},\ and\ \citenamefont
  {Wieck}}]{PhysRevB.87.205203}%
  \BibitemOpen
  \bibfield  {author} {\bibinfo {author} {\bibfnamefont {J.-H.}\ \bibnamefont
  {Quast}}, \bibinfo {author} {\bibfnamefont {T.}~\bibnamefont {Henn}},
  \bibinfo {author} {\bibfnamefont {T.}~\bibnamefont {Kiessling}}, \bibinfo
  {author} {\bibfnamefont {W.}~\bibnamefont {Ossau}}, \bibinfo {author}
  {\bibfnamefont {L.~W.}\ \bibnamefont {Molenkamp}}, \bibinfo {author}
  {\bibfnamefont {D.}~\bibnamefont {Reuter}}, \ and\ \bibinfo {author}
  {\bibfnamefont {A.~D.}\ \bibnamefont {Wieck}},\ }\href {\doibase
  10.1103/PhysRevB.87.205203} {\bibfield  {journal} {\bibinfo  {journal} {Phys.
  Rev. B}\ }\textbf {\bibinfo {volume} {87}},\ \bibinfo {pages} {205203}
  (\bibinfo {year} {2013})}\BibitemShut {NoStop}%
\bibitem [{\citenamefont {Weber}\ \emph {et~al.}(2011)\citenamefont {Weber},
  \citenamefont {Benko},\ and\ \citenamefont {Hiew}}]{Weber_JAP109_106101_11}%
  \BibitemOpen
  \bibfield  {author} {\bibinfo {author} {\bibfnamefont {C.~P.}\ \bibnamefont
  {Weber}}, \bibinfo {author} {\bibfnamefont {C.~A.}\ \bibnamefont {Benko}}, \
  and\ \bibinfo {author} {\bibfnamefont {S.~C.}\ \bibnamefont {Hiew}},\ }\href
  {\doibase 10.1063/1.3592272} {\bibfield  {journal} {\bibinfo  {journal} {J.
  App. Phys.}\ }\textbf {\bibinfo {volume} {109}},\ \bibinfo {pages} {106101}
  (\bibinfo {year} {2011})}\BibitemShut {NoStop}%
\bibitem [{\citenamefont {Henn}\ \emph {et~al.}(2013)\citenamefont {Henn},
  \citenamefont {Kiessling}, \citenamefont {Ossau}, \citenamefont {Molenkamp},
  \citenamefont {Reuter},\ and\ \citenamefont {Wieck}}]{PhysRevB.88.195202}%
  \BibitemOpen
  \bibfield  {author} {\bibinfo {author} {\bibfnamefont {T.}~\bibnamefont
  {Henn}}, \bibinfo {author} {\bibfnamefont {T.}~\bibnamefont {Kiessling}},
  \bibinfo {author} {\bibfnamefont {W.}~\bibnamefont {Ossau}}, \bibinfo
  {author} {\bibfnamefont {L.~W.}\ \bibnamefont {Molenkamp}}, \bibinfo {author}
  {\bibfnamefont {D.}~\bibnamefont {Reuter}}, \ and\ \bibinfo {author}
  {\bibfnamefont {A.~D.}\ \bibnamefont {Wieck}},\ }\href {\doibase
  10.1103/PhysRevB.88.195202} {\bibfield  {journal} {\bibinfo  {journal} {Phys.
  Rev. B}\ }\textbf {\bibinfo {volume} {88}},\ \bibinfo {pages} {195202}
  (\bibinfo {year} {2013})}\BibitemShut {NoStop}%
\bibitem [{\citenamefont {Cameron}\ \emph {et~al.}(1996)\citenamefont
  {Cameron}, \citenamefont {Riblet},\ and\ \citenamefont
  {Miller}}]{Cameron_PRL76_4793_96}%
  \BibitemOpen
  \bibfield  {author} {\bibinfo {author} {\bibfnamefont {A.~R.}\ \bibnamefont
  {Cameron}}, \bibinfo {author} {\bibfnamefont {P.}~\bibnamefont {Riblet}}, \
  and\ \bibinfo {author} {\bibfnamefont {A.}~\bibnamefont {Miller}},\
  }\href@noop {} {\bibfield  {journal} {\bibinfo  {journal} {Phys. Rev. Lett.}\
  }\textbf {\bibinfo {volume} {76}},\ \bibinfo {pages} {4793} (\bibinfo {year}
  {1996})}\BibitemShut {NoStop}%
\bibitem [{\citenamefont {Eldridge}\ \emph {et~al.}(2008)\citenamefont
  {Eldridge}, \citenamefont {Leyland}, \citenamefont {Lagoudakis},
  \citenamefont {Karimov}, \citenamefont {Henini}, \citenamefont {Taylor},
  \citenamefont {Phillips},\ and\ \citenamefont
  {Harley}}]{Eldridge_PRB77_125344_08}%
  \BibitemOpen
  \bibfield  {author} {\bibinfo {author} {\bibfnamefont {P.~S.}\ \bibnamefont
  {Eldridge}}, \bibinfo {author} {\bibfnamefont {W.~J.~H.}\ \bibnamefont
  {Leyland}}, \bibinfo {author} {\bibfnamefont {P.~G.}\ \bibnamefont
  {Lagoudakis}}, \bibinfo {author} {\bibfnamefont {O.~Z.}\ \bibnamefont
  {Karimov}}, \bibinfo {author} {\bibfnamefont {M.}~\bibnamefont {Henini}},
  \bibinfo {author} {\bibfnamefont {D.}~\bibnamefont {Taylor}}, \bibinfo
  {author} {\bibfnamefont {R.~T.}\ \bibnamefont {Phillips}}, \ and\ \bibinfo
  {author} {\bibfnamefont {R.~T.}\ \bibnamefont {Harley}},\ }\href {\doibase
  10.1103/PhysRevB.77.125344} {\bibfield  {journal} {\bibinfo  {journal} {Phys.
  Rev. B}\ }\textbf {\bibinfo {volume} {77}},\ \bibinfo {pages} {125344}
  (\bibinfo {year} {2008})}\BibitemShut {NoStop}%
\bibitem [{\citenamefont {Zhao}\ \emph {et~al.}(2009)\citenamefont {Zhao},
  \citenamefont {Mower},\ and\ \citenamefont {Vignale}}]{Zhao_PRB79_115321_09}%
  \BibitemOpen
  \bibfield  {author} {\bibinfo {author} {\bibfnamefont {H.}~\bibnamefont
  {Zhao}}, \bibinfo {author} {\bibfnamefont {M.}~\bibnamefont {Mower}}, \ and\
  \bibinfo {author} {\bibfnamefont {G.}~\bibnamefont {Vignale}},\ }\href
  {\doibase 10.1103/PhysRevB.79.115321} {\bibfield  {journal} {\bibinfo
  {journal} {Phys. Rev. B}\ }\textbf {\bibinfo {volume} {79}},\ \bibinfo
  {pages} {115321} (\bibinfo {year} {2009})}\BibitemShut {NoStop}%
\bibitem [{\citenamefont {Hu}\ \emph {et~al.}(2011)\citenamefont {Hu},
  \citenamefont {Ye}, \citenamefont {Wang}, \citenamefont {Tian}, \citenamefont
  {Wang}, \citenamefont {Wang}, \citenamefont {Liu},\ and\ \citenamefont
  {Marie}}]{Hu_NRL6_149_11}%
  \BibitemOpen
  \bibfield  {author} {\bibinfo {author} {\bibfnamefont {C.}~\bibnamefont
  {Hu}}, \bibinfo {author} {\bibfnamefont {H.}~\bibnamefont {Ye}}, \bibinfo
  {author} {\bibfnamefont {G.}~\bibnamefont {Wang}}, \bibinfo {author}
  {\bibfnamefont {H.}~\bibnamefont {Tian}}, \bibinfo {author} {\bibfnamefont
  {W.}~\bibnamefont {Wang}}, \bibinfo {author} {\bibfnamefont {W.}~\bibnamefont
  {Wang}}, \bibinfo {author} {\bibfnamefont {B.}~\bibnamefont {Liu}}, \ and\
  \bibinfo {author} {\bibfnamefont {X.}~\bibnamefont {Marie}},\ }\href
  {\doibase 10.1186/1556-276X-6-149} {\bibfield  {journal} {\bibinfo  {journal}
  {Nanoscale Res. Lett.}\ }\textbf {\bibinfo {volume} {6}},\ \bibinfo {pages}
  {149} (\bibinfo {year} {2011})}\BibitemShut {NoStop}%
\bibitem [{\citenamefont {Henn}\ \emph {et~al.}(2014)\citenamefont {Henn},
  \citenamefont {Quast}, \citenamefont {Kiessling}, \citenamefont {Ossau},
  \citenamefont {Molenkamp}, \citenamefont {Reuter}, \citenamefont {Wieck},
  \citenamefont {Biermann},\ and\ \citenamefont {Santos}}]{PVS276}%
  \BibitemOpen
  \bibfield  {author} {\bibinfo {author} {\bibfnamefont {T.}~\bibnamefont
  {Henn}}, \bibinfo {author} {\bibfnamefont {J.-H.}\ \bibnamefont {Quast}},
  \bibinfo {author} {\bibfnamefont {T.}~\bibnamefont {Kiessling}}, \bibinfo
  {author} {\bibfnamefont {W.}~\bibnamefont {Ossau}}, \bibinfo {author}
  {\bibfnamefont {L.~W.}\ \bibnamefont {Molenkamp}}, \bibinfo {author}
  {\bibfnamefont {D.}~\bibnamefont {Reuter}}, \bibinfo {author} {\bibfnamefont
  {A.~D.}\ \bibnamefont {Wieck}}, \bibinfo {author} {\bibfnamefont
  {K.}~\bibnamefont {Biermann}}, \ and\ \bibinfo {author} {\bibfnamefont
  {P.~V.}\ \bibnamefont {Santos}},\ }\href {\doibase 10.1002/pssb.201350192}
  {\bibfield  {journal} {\bibinfo  {journal} {Phys. Status Solidi B}\ }\textbf
  {\bibinfo {volume} {251}},\ \bibinfo {pages} {1839} (\bibinfo {year}
  {2014})}\BibitemShut {NoStop}%
\bibitem [{\citenamefont {Weber}\ \emph {et~al.}(2005)\citenamefont {Weber},
  \citenamefont {Gedik}, \citenamefont {Moore}, \citenamefont {Orenstein},
  \citenamefont {Stephens},\ and\ \citenamefont
  {Awschalom}}]{Weber_N437_1330_05}%
  \BibitemOpen
  \bibfield  {author} {\bibinfo {author} {\bibfnamefont {C.~P.}\ \bibnamefont
  {Weber}}, \bibinfo {author} {\bibfnamefont {N.}~\bibnamefont {Gedik}},
  \bibinfo {author} {\bibfnamefont {J.~E.}\ \bibnamefont {Moore}}, \bibinfo
  {author} {\bibfnamefont {J.}~\bibnamefont {Orenstein}}, \bibinfo {author}
  {\bibfnamefont {J.}~\bibnamefont {Stephens}}, \ and\ \bibinfo {author}
  {\bibfnamefont {D.~D.}\ \bibnamefont {Awschalom}},\ }\href {\doibase
  10.1038/nature04206} {\bibfield  {journal} {\bibinfo  {journal} {Nature}\
  }\textbf {\bibinfo {volume} {437}},\ \bibinfo {pages} {1330} (\bibinfo {year}
  {2005})}\BibitemShut {NoStop}%
\bibitem [{\citenamefont {Carter}\ \emph {et~al.}(2006)\citenamefont {Carter},
  \citenamefont {Chen},\ and\ \citenamefont
  {Cundiff}}]{Carter_PRL97_136602_06}%
  \BibitemOpen
  \bibfield  {author} {\bibinfo {author} {\bibfnamefont {S.~G.}\ \bibnamefont
  {Carter}}, \bibinfo {author} {\bibfnamefont {Z.}~\bibnamefont {Chen}}, \ and\
  \bibinfo {author} {\bibfnamefont {S.~T.}\ \bibnamefont {Cundiff}},\
  }\href@noop {} {\bibfield  {journal} {\bibinfo  {journal} {Phys. Rev. Lett.}\
  }\textbf {\bibinfo {volume} {97}},\ \bibinfo {pages} {136602} (\bibinfo
  {year} {2006})}\BibitemShut {NoStop}%
\bibitem [{\citenamefont {V\"olkl}\ \emph {et~al.}(2011)\citenamefont
  {V\"olkl}, \citenamefont {Griesbeck}, \citenamefont {Tarasenko},
  \citenamefont {Schuh}, \citenamefont {Wegscheider}, \citenamefont
  {Sch\"uller},\ and\ \citenamefont {Korn}}]{Volkl_PRB83_241306_11}%
  \BibitemOpen
  \bibfield  {author} {\bibinfo {author} {\bibfnamefont {R.}~\bibnamefont
  {V\"olkl}}, \bibinfo {author} {\bibfnamefont {M.}~\bibnamefont {Griesbeck}},
  \bibinfo {author} {\bibfnamefont {S.~A.}\ \bibnamefont {Tarasenko}}, \bibinfo
  {author} {\bibfnamefont {D.}~\bibnamefont {Schuh}}, \bibinfo {author}
  {\bibfnamefont {W.}~\bibnamefont {Wegscheider}}, \bibinfo {author}
  {\bibfnamefont {C.}~\bibnamefont {Sch\"uller}}, \ and\ \bibinfo {author}
  {\bibfnamefont {T.}~\bibnamefont {Korn}},\ }\href {\doibase
  10.1103/PhysRevB.83.241306} {\bibfield  {journal} {\bibinfo  {journal} {Phys.
  Rev. B}\ }\textbf {\bibinfo {volume} {83}},\ \bibinfo {pages} {241306}
  (\bibinfo {year} {2011})}\BibitemShut {NoStop}%
\bibitem [{\citenamefont {V\"olkl}\ \emph {et~al.}(2014)\citenamefont
  {V\"olkl}, \citenamefont {Schwemmer}, \citenamefont {Griesbeck},
  \citenamefont {Tarasenko}, \citenamefont {Schuh}, \citenamefont
  {Wegscheider}, \citenamefont {Sch\"uller},\ and\ \citenamefont
  {Korn}}]{Volkl_PRB89_75424_14}%
  \BibitemOpen
  \bibfield  {author} {\bibinfo {author} {\bibfnamefont {R.}~\bibnamefont
  {V\"olkl}}, \bibinfo {author} {\bibfnamefont {M.}~\bibnamefont {Schwemmer}},
  \bibinfo {author} {\bibfnamefont {M.}~\bibnamefont {Griesbeck}}, \bibinfo
  {author} {\bibfnamefont {S.~A.}\ \bibnamefont {Tarasenko}}, \bibinfo {author}
  {\bibfnamefont {D.}~\bibnamefont {Schuh}}, \bibinfo {author} {\bibfnamefont
  {W.}~\bibnamefont {Wegscheider}}, \bibinfo {author} {\bibfnamefont
  {C.}~\bibnamefont {Sch\"uller}}, \ and\ \bibinfo {author} {\bibfnamefont
  {T.}~\bibnamefont {Korn}},\ }\href {\doibase 10.1103/PhysRevB.89.075424}
  {\bibfield  {journal} {\bibinfo  {journal} {Phys. Rev. B}\ }\textbf {\bibinfo
  {volume} {89}},\ \bibinfo {pages} {075424} (\bibinfo {year}
  {2014})}\BibitemShut {NoStop}%
\bibitem [{\citenamefont {Altmann}\ \emph {et~al.}(2014)\citenamefont
  {Altmann}, \citenamefont {Walser}, \citenamefont {Reichl}, \citenamefont
  {Wegscheider},\ and\ \citenamefont {Salis}}]{Altmann_PRB90_201306_14}%
  \BibitemOpen
  \bibfield  {author} {\bibinfo {author} {\bibfnamefont {P.}~\bibnamefont
  {Altmann}}, \bibinfo {author} {\bibfnamefont {M.~P.}\ \bibnamefont {Walser}},
  \bibinfo {author} {\bibfnamefont {C.}~\bibnamefont {Reichl}}, \bibinfo
  {author} {\bibfnamefont {W.}~\bibnamefont {Wegscheider}}, \ and\ \bibinfo
  {author} {\bibfnamefont {G.}~\bibnamefont {Salis}},\ }\href {\doibase
  10.1103/PhysRevB.90.201306} {\bibfield  {journal} {\bibinfo  {journal} {Phys.
  Rev. B}\ }\textbf {\bibinfo {volume} {90}},\ \bibinfo {pages} {201306(R)}
  (\bibinfo {year} {2014})}\BibitemShut {NoStop}%
\bibitem [{\citenamefont {Kohda}\ \emph {et~al.}(2015)\citenamefont {Kohda},
  \citenamefont {Altmann}, \citenamefont {Schuh}, \citenamefont {Ganichev},
  \citenamefont {Wegscheider},\ and\ \citenamefont
  {Salis}}]{Kohda_APL107_172402_15}%
  \BibitemOpen
  \bibfield  {author} {\bibinfo {author} {\bibfnamefont {M.}~\bibnamefont
  {Kohda}}, \bibinfo {author} {\bibfnamefont {P.}~\bibnamefont {Altmann}},
  \bibinfo {author} {\bibfnamefont {D.}~\bibnamefont {Schuh}}, \bibinfo
  {author} {\bibfnamefont {S.~D.}\ \bibnamefont {Ganichev}}, \bibinfo {author}
  {\bibfnamefont {W.}~\bibnamefont {Wegscheider}}, \ and\ \bibinfo {author}
  {\bibnamefont {Salis}},\ }\href {\doibase 10.1063/1.4934671} {\bibfield
  {journal} {\bibinfo  {journal} {App. Phys. Lett.}\ }\textbf {\bibinfo
  {volume} {107}},\ \bibinfo {pages} {172402} (\bibinfo {year}
  {2015})}\BibitemShut {NoStop}%
\bibitem [{\citenamefont {Altmann}\ \emph {et~al.}(2015)\citenamefont
  {Altmann}, \citenamefont {Kohda}, \citenamefont {Reichl}, \citenamefont
  {Wegscheider},\ and\ \citenamefont {Salis}}]{Altmann_PRB92_235304_15}%
  \BibitemOpen
  \bibfield  {author} {\bibinfo {author} {\bibfnamefont {P.}~\bibnamefont
  {Altmann}}, \bibinfo {author} {\bibfnamefont {M.}~\bibnamefont {Kohda}},
  \bibinfo {author} {\bibfnamefont {C.}~\bibnamefont {Reichl}}, \bibinfo
  {author} {\bibfnamefont {W.}~\bibnamefont {Wegscheider}}, \ and\ \bibinfo
  {author} {\bibfnamefont {G.}~\bibnamefont {Salis}},\ }\href {\doibase
  10.1103/PhysRevB.92.235304} {\bibfield  {journal} {\bibinfo  {journal} {Phys.
  Rev. B}\ }\textbf {\bibinfo {volume} {92}},\ \bibinfo {pages} {235304}
  (\bibinfo {year} {2015})}\BibitemShut {NoStop}%
\bibitem [{\citenamefont {D'Amico}\ and\ \citenamefont
  {Vignale}(2001)}]{amico01a}%
  \BibitemOpen
  \bibfield  {author} {\bibinfo {author} {\bibfnamefont {I.}~\bibnamefont
  {D'Amico}}\ and\ \bibinfo {author} {\bibfnamefont {G.}~\bibnamefont
  {Vignale}},\ }\href@noop {} {\bibfield  {journal} {\bibinfo  {journal}
  {Europhys. Lett.}\ }\textbf {\bibinfo {volume} {55}},\ \bibinfo {pages} {566}
  (\bibinfo {year} {2001})}\BibitemShut {NoStop}%
\bibitem [{\citenamefont {Wang}\ \emph
  {et~al.}(2013{\natexlab{b}})\citenamefont {Wang}, \citenamefont {Liu},
  \citenamefont {Balocchi}, \citenamefont {Renucci}, \citenamefont {Zhu},
  \citenamefont {Amand}, \citenamefont {Fontaine},\ and\ \citenamefont
  {Marie}}]{Wang_NC4_2372_13}%
  \BibitemOpen
  \bibfield  {author} {\bibinfo {author} {\bibfnamefont {G.}~\bibnamefont
  {Wang}}, \bibinfo {author} {\bibfnamefont {B.~L.}\ \bibnamefont {Liu}},
  \bibinfo {author} {\bibfnamefont {A.}~\bibnamefont {Balocchi}}, \bibinfo
  {author} {\bibfnamefont {P.}~\bibnamefont {Renucci}}, \bibinfo {author}
  {\bibfnamefont {C.~R.}\ \bibnamefont {Zhu}}, \bibinfo {author} {\bibfnamefont
  {T.}~\bibnamefont {Amand}}, \bibinfo {author} {\bibfnamefont
  {C.}~\bibnamefont {Fontaine}}, \ and\ \bibinfo {author} {\bibfnamefont
  {X.}~\bibnamefont {Marie}},\ }\href {\doibase 10.1038/ncomms3372} {\bibfield
  {journal} {\bibinfo  {journal} {Nature Com}\ }\textbf {\bibinfo {volume}
  {4}},\ \bibinfo {pages} {2372} (\bibinfo {year}
  {2013}{\natexlab{b}})}\BibitemShut {NoStop}%
\bibitem [{\citenamefont {Damen}\ \emph {et~al.}(1991)\citenamefont {Damen},
  \citenamefont {Vi\~na}, \citenamefont {Cunningham}, \citenamefont {Shah},\
  and\ \citenamefont {Sham}}]{PhysRevLett.67.3432}%
  \BibitemOpen
  \bibfield  {author} {\bibinfo {author} {\bibfnamefont {T.~C.}\ \bibnamefont
  {Damen}}, \bibinfo {author} {\bibfnamefont {L.}~\bibnamefont {Vi\~na}},
  \bibinfo {author} {\bibfnamefont {J.~E.}\ \bibnamefont {Cunningham}},
  \bibinfo {author} {\bibfnamefont {J.}~\bibnamefont {Shah}}, \ and\ \bibinfo
  {author} {\bibfnamefont {L.~J.}\ \bibnamefont {Sham}},\ }\href {\doibase
  10.1103/PhysRevLett.67.3432} {\bibfield  {journal} {\bibinfo  {journal}
  {Phys. Rev. Lett.}\ }\textbf {\bibinfo {volume} {67}},\ \bibinfo {pages}
  {3432} (\bibinfo {year} {1991})}\BibitemShut {NoStop}%
\bibitem [{\citenamefont {Hilton}\ and\ \citenamefont
  {Tang}(2002)}]{PhysRevLett.89.146601}%
  \BibitemOpen
  \bibfield  {author} {\bibinfo {author} {\bibfnamefont {D.~J.}\ \bibnamefont
  {Hilton}}\ and\ \bibinfo {author} {\bibfnamefont {C.~L.}\ \bibnamefont
  {Tang}},\ }\href {\doibase 10.1103/PhysRevLett.89.146601} {\bibfield
  {journal} {\bibinfo  {journal} {Phys. Rev. Lett.}\ }\textbf {\bibinfo
  {volume} {89}},\ \bibinfo {pages} {146601} (\bibinfo {year}
  {2002})}\BibitemShut {NoStop}%
\bibitem [{\citenamefont {Remeika}\ \emph {et~al.}(2009)\citenamefont
  {Remeika}, \citenamefont {Graves}, \citenamefont {Hammack}, \citenamefont
  {Meyertholen}, \citenamefont {Fogler}, \citenamefont {Butov}, \citenamefont
  {Hanson},\ and\ \citenamefont {Gossard}}]{Remeika_PRL102_186803_09}%
  \BibitemOpen
  \bibfield  {author} {\bibinfo {author} {\bibfnamefont {M.}~\bibnamefont
  {Remeika}}, \bibinfo {author} {\bibfnamefont {J.~C.}\ \bibnamefont {Graves}},
  \bibinfo {author} {\bibfnamefont {A.~T.}\ \bibnamefont {Hammack}}, \bibinfo
  {author} {\bibfnamefont {A.~D.}\ \bibnamefont {Meyertholen}}, \bibinfo
  {author} {\bibfnamefont {M.~M.}\ \bibnamefont {Fogler}}, \bibinfo {author}
  {\bibfnamefont {L.}~\bibnamefont {Butov}}, \bibinfo {author} {\bibfnamefont
  {M.}~\bibnamefont {Hanson}}, \ and\ \bibinfo {author} {\bibfnamefont {A.~C.}\
  \bibnamefont {Gossard}},\ }\href {\doibase 10.1103/PhysRevLett.102.186803}
  {\bibfield  {journal} {\bibinfo  {journal} {Phys. Rev. Lett.}\ }\textbf
  {\bibinfo {volume} {102}},\ \bibinfo {pages} {186803} (\bibinfo {year}
  {2009})}\BibitemShut {NoStop}%
\bibitem [{\citenamefont {Meier}\ and\ \citenamefont
  {Zakharchenya}(1984)}]{Meier_84}%
  \BibitemOpen
  \bibfield  {author} {\bibinfo {author} {\bibfnamefont {F.}~\bibnamefont
  {Meier}}\ and\ \bibinfo {author} {\bibfnamefont {B.~P.}\ \bibnamefont
  {Zakharchenya}},\ }\href@noop {} {\emph {\bibinfo {title} {Optical
  orientation}}},\ edited by\ \bibinfo {editor} {\bibfnamefont {V.~M.}\
  \bibnamefont {Agranovich}}\ and\ \bibinfo {editor} {\bibfnamefont {A.~A.}\
  \bibnamefont {Maradudin}},\ \bibinfo {series} {Modern problems in condensed
  matter physics}\ No.~\bibinfo {number} {8}\ (\bibinfo  {publisher}
  {North-Holland},\ \bibinfo {address} {Amsterdam, The Netherlands},\ \bibinfo
  {year} {1984})\BibitemShut {NoStop}%
\bibitem [{\citenamefont {Froltsov}(2001)}]{Froltsov_PRB64_45311_01}%
  \BibitemOpen
  \bibfield  {author} {\bibinfo {author} {\bibfnamefont {V.~A.}\ \bibnamefont
  {Froltsov}},\ }\href {\doibase 10.1103/PhysRevB.64.045311} {\bibfield
  {journal} {\bibinfo  {journal} {Phys. Rev. B}\ }\textbf {\bibinfo {volume}
  {64}},\ \bibinfo {pages} {045311} (\bibinfo {year} {2001})}\BibitemShut
  {NoStop}%
\bibitem [{\citenamefont {Stanescu}\ and\ \citenamefont
  {Galitski}(2007)}]{Stanescu_PRB75_125307_07}%
  \BibitemOpen
  \bibfield  {author} {\bibinfo {author} {\bibfnamefont {T.~D.}\ \bibnamefont
  {Stanescu}}\ and\ \bibinfo {author} {\bibfnamefont {V.}~\bibnamefont
  {Galitski}},\ }\href@noop {} {\bibfield  {journal} {\bibinfo  {journal}
  {Phys. Rev. B}\ }\textbf {\bibinfo {volume} {75}},\ \bibinfo {pages} {125307}
  (\bibinfo {year} {2007})}\BibitemShut {NoStop}%
\bibitem [{\citenamefont {Yang}\ \emph {et~al.}(2010)\citenamefont {Yang},
  \citenamefont {Orenstein},\ and\ \citenamefont {Lee}}]{Yang_PRB82_155324_10}%
  \BibitemOpen
  \bibfield  {author} {\bibinfo {author} {\bibfnamefont {L.}~\bibnamefont
  {Yang}}, \bibinfo {author} {\bibfnamefont {J.}~\bibnamefont {Orenstein}}, \
  and\ \bibinfo {author} {\bibfnamefont {D.-H.}\ \bibnamefont {Lee}},\ }\href
  {\doibase 10.1103/PhysRevB.82.155324} {\bibfield  {journal} {\bibinfo
  {journal} {Phys. Rev. B}\ }\textbf {\bibinfo {volume} {82}},\ \bibinfo
  {pages} {155324} (\bibinfo {year} {2010})}\BibitemShut {NoStop}%
\bibitem [{\citenamefont {Poshakinskiy}\ and\ \citenamefont
  {Tarasenko}(2015)}]{Poshakinskiy_PRB92_45308_15}%
  \BibitemOpen
  \bibfield  {author} {\bibinfo {author} {\bibfnamefont {A.~V.}\ \bibnamefont
  {Poshakinskiy}}\ and\ \bibinfo {author} {\bibfnamefont {S.~A.}\ \bibnamefont
  {Tarasenko}},\ }\href {\doibase 10.1103/PhysRevB.92.045308} {\bibfield
  {journal} {\bibinfo  {journal} {Phys. Rev. B}\ }\textbf {\bibinfo {volume}
  {92}},\ \bibinfo {pages} {045308} (\bibinfo {year} {2015})}\BibitemShut
  {NoStop}%
\bibitem [{\citenamefont {Bernevig}\ \emph {et~al.}(2006)\citenamefont
  {Bernevig}, \citenamefont {Orenstein},\ and\ \citenamefont
  {Zhang}}]{Bernevig_PRL97_236601_06}%
  \BibitemOpen
  \bibfield  {author} {\bibinfo {author} {\bibfnamefont {B.~A.}\ \bibnamefont
  {Bernevig}}, \bibinfo {author} {\bibfnamefont {J.}~\bibnamefont {Orenstein}},
  \ and\ \bibinfo {author} {\bibfnamefont {S.-C.}\ \bibnamefont {Zhang}},\
  }\href@noop {} {\bibfield  {journal} {\bibinfo  {journal} {Phys. Rev. Lett.}\
  }\textbf {\bibinfo {volume} {97}},\ \bibinfo {pages} {236601} (\bibinfo
  {year} {2006})}\BibitemShut {NoStop}%
\bibitem [{\citenamefont {Koralek}\ \emph {et~al.}(2009)\citenamefont
  {Koralek}, \citenamefont {Weber}, \citenamefont {Orenstein}, \citenamefont
  {Bernevig}, \citenamefont {Zhang}, \citenamefont {Mack},\ and\ \citenamefont
  {Awschalom}}]{Koralek_N458_610_09}%
  \BibitemOpen
  \bibfield  {author} {\bibinfo {author} {\bibfnamefont {J.~D.}\ \bibnamefont
  {Koralek}}, \bibinfo {author} {\bibfnamefont {C.~P.}\ \bibnamefont {Weber}},
  \bibinfo {author} {\bibfnamefont {J.}~\bibnamefont {Orenstein}}, \bibinfo
  {author} {\bibfnamefont {B.~A.}\ \bibnamefont {Bernevig}}, \bibinfo {author}
  {\bibfnamefont {S.-C.}\ \bibnamefont {Zhang}}, \bibinfo {author}
  {\bibfnamefont {S.}~\bibnamefont {Mack}}, \ and\ \bibinfo {author}
  {\bibfnamefont {D.~D.}\ \bibnamefont {Awschalom}},\ }\href
  {http://dx.doi.org/10.1038/nature07871} {\bibfield  {journal} {\bibinfo
  {journal} {Nature}\ }\textbf {\bibinfo {volume} {458}},\ \bibinfo {pages}
  {610} (\bibinfo {year} {2009})}\BibitemShut {NoStop}%
\bibitem [{\citenamefont {Walser}\ \emph {et~al.}(2012)\citenamefont {Walser},
  \citenamefont {Reichl}, \citenamefont {Wegscheider},\ and\ \citenamefont
  {Salis}}]{Walser2012}%
  \BibitemOpen
  \bibfield  {author} {\bibinfo {author} {\bibfnamefont {M.~P.}\ \bibnamefont
  {Walser}}, \bibinfo {author} {\bibfnamefont {C.}~\bibnamefont {Reichl}},
  \bibinfo {author} {\bibfnamefont {W.}~\bibnamefont {Wegscheider}}, \ and\
  \bibinfo {author} {\bibfnamefont {G.}~\bibnamefont {Salis}},\ }\href
  {http://dx.doi.org/10.1038/nphys2383} {\bibfield  {journal} {\bibinfo
  {journal} {Nat. Phys.}\ }\textbf {\bibinfo {volume} {8}},\ \bibinfo {pages}
  {757} (\bibinfo {year} {2012})}\BibitemShut {NoStop}%
\bibitem [{Note1()}]{Note1}%
  \BibitemOpen
  \bibinfo {note} {Although a Bessel function is a more appropriate solution
  for a radially symmetric differential equation than a harmonic oscillation,
  both functions behave similar for $r\ll r_n$.}\BibitemShut {Stop}%
\bibitem [{Note2()}]{Note2}%
  \BibitemOpen
  \bibinfo {note} {The repulsive forces at $r=0$ must be $F_r=0$ because of the
  radial symmetry of the carrier density. Due its gaussian shape, the carrier
  concentration is the largest at the center and decays radially towards zero.
  Therefore, $F_r$ must initially increase with $r$ until reaching a maximum
  around $w_n/2$, and then decay towards $F_r\rightarrow 0$.}\BibitemShut
  {Stop}%
\bibitem [{\citenamefont {Salis}\ \emph {et~al.}(2014)\citenamefont {Salis},
  \citenamefont {Walser}, \citenamefont {Altmann}, \citenamefont {Reichl},\
  and\ \citenamefont {Wegscheider}}]{Salis_PRB89_45304_14}%
  \BibitemOpen
  \bibfield  {author} {\bibinfo {author} {\bibfnamefont {G.}~\bibnamefont
  {Salis}}, \bibinfo {author} {\bibfnamefont {M.~P.}\ \bibnamefont {Walser}},
  \bibinfo {author} {\bibfnamefont {P.}~\bibnamefont {Altmann}}, \bibinfo
  {author} {\bibfnamefont {C.}~\bibnamefont {Reichl}}, \ and\ \bibinfo {author}
  {\bibfnamefont {W.}~\bibnamefont {Wegscheider}},\ }\href {\doibase
  10.1103/PhysRevB.89.045304} {\bibfield  {journal} {\bibinfo  {journal} {Phys.
  Rev. B}\ }\textbf {\bibinfo {volume} {89}},\ \bibinfo {pages} {045304}
  (\bibinfo {year} {2014})}\BibitemShut {NoStop}%
\end{thebibliography}


\end{document}